\documentclass{JFM-FLM_Au}

\usepackage{amsmath}
\usepackage{bm}
\usepackage{xcolor}

\newcommand{\dd}{\mathrm{d}}
\newcommand{\ii}{\mathrm{i}}
\newcommand{\bu}{\bm{u}}
\newcommand{\bx}{\bm{x}}
\newcommand{\bk}{\bm{k}}

\newcommand{\cJ}{\mathcal{J}}
\newcommand{\cB}{\mathcal{B}}

\newcommand{\cN}{\mathcal{N}}

\newcommand{\cL}{\mathcal{L}}
\newcommand{\cU}{\mathcal{U}}
\newcommand{\cP}{\mathcal{P}}
\newcommand{\cX}{\mathcal{X}}
\newcommand{\Jt}{\cJ_{\mathrm{t}}}

\newcommand{\Jc}{\cJ_{\mathrm{conv}}}
\newcommand{\Jx}{\cJ_{\mathrm{x}}}
\newcommand{\Jxo}{\cJ_{\partial x}}
\newcommand{\tildeJxo}{\tilde{\cJ}_{\partial x}}

\hypersetup{linkcolor=black, urlcolor=black}

\lefttitle{D. Lasagna and T. Burton}
\righttitle{Journal of Fluid Mechanics}

\title{Variational Newton--Krylov descent for periodic orbits of two-dimensional Kolmogorov flow}

\author{Davide Lasagna\aff{1} and Thomas Burton\aff{1}}

\affiliation{\aff{1}Faculty of Engineering and Physical Sciences, University of Southampton, Southampton SO17 1BJ, UK}

\corresau{Davide Lasagna, \email{davide.lasagna@soton.ac.uk}}

\begin{document}

\maketitle

\begin{abstract}
A matrix-free Newton--Krylov method for computing long unstable periodic orbits of turbulent flows is presented. The method arises from the space-time variational formulation of Lan \& Cvitanovi\'c (\textit{Phys. Rev. E}, vol. 69, 2004, 016217), and runs on graphics processing units with a preconditioner from time-spectral methods that retains advection. In two-dimensional Kolmogorov flow at $\Rey = 40$, near recurrences remain about as frequent at long periods as at moderate ones, and searches from them converge to 183 distinct solutions with periods up to 191, about forty escape times, all but one verified by multiple shooting. The primary question is then whether such a long orbit can be taken, on its own, as a proxy for the attractor, and thus capture the statistics of the turbulent flow. The mean dissipation and energy of the long orbits lie close to the medians of the flow, and the distribution of the dissipation along a long orbit is comparable with that of a turbulent segment of the same duration, except in its tail, which most orbits miss. Their leading Floquet exponents appear to lie slightly below the Lyapunov spectrum of the flow, as may be expected of quieter orbits. The typical distance of the turbulent flow from a long orbit decreases as the period grows, but with an unfavourable scaling set by the correlation dimension of the attractor, which suggests that ever longer orbits may be needed as the Reynolds number grows.
\end{abstract}


\section{Introduction}
\label{sec:introduction}

Over the past two decades a view of turbulence has emerged in which a turbulent trajectory shadows a skeleton of unstable invariant solutions of the Navier--Stokes equations, repelled from one towards others along their unstable manifolds \citep{kawahara2012significance, graham2021exact}. Unstable periodic orbits, which capture the cycles of growth and breakdown of coherent structures, play the central role in this skeleton, and since the first was found \citep{kawahara2001periodic} many of relatively short period have been computed in canonical flows \citep{viswanath2007recurrent, chandler2013invariant, budanur2017relative}. The main motivation for computing them is the prospect of a deterministic account of turbulent statistics. Periodic orbit theory expresses long-time averages as weighted sums over such orbits \citep{artuso1990recyclingI,artuso1990recyclingII,chaosbook}, but its cycle expansions converge only once all the short orbits up to a given period are known, a requirement out of reach in turbulent flows of many degrees of freedom, although a complete set has recently been computed in a minimal Taylor--Couette flow \citep{wang2025mathematically}. Libraries of a few tens to hundreds of short orbits therefore approximate the density of an observable by a weighted combination of those of the individual orbits, with equal weights or weights from periodic orbit theory \citep{chandler2013invariant, lucas2015recurrent} or, more recently, with data-driven weights obtained a posteriori from how often the turbulent flow visits each orbit \citep{page2024recurrent} or from a fit, by minimising the Kullback--Leibler divergence between the reconstructed and the turbulent densities \citep{redfern2024dynamically, cleary2025dynamical}. The recent libraries of two-dimensional Kolmogorov flow, of a hundred or more mostly short orbits, reproduce the densities of the dissipation and of the energy closely, with weights that decrease with the instability of the orbits \citep{redfern2024dynamically}. Since each short orbit covers only a narrow range of the dissipation, these libraries need bursting orbits to reach the tail, and searches have targeted them explicitly \citep{page2024exact, page2024recurrent}.

Such reconstructions are a valuable test of the theory, but our focus is on a different question, one that relates to many problems in flow control, optimisation and system identification. A turbulence simulation does not tell us directly how the means change with the parameters of the flow. This sensitivity is often obtained with adjoint-based approaches when many parameters are involved, but on an open chaotic trajectory the adjoint solution grows exponentially \citep{lea2000sensitivity}. Libraries do not give this sensitivity directly either. Their weights are obtained at one set of parameters, and the reconstructions deteriorate when the orbits are continued in the Reynolds number \citep{cleary2025dynamical}. For this purpose we proposed to solve the adjoint problem on periodic orbits \citep{lasagna2018sensitivity}, where the periodicity condition removes the exponential growth, despite the instability, and gives a well-conditioned sensitivity of the mean over the orbit \citep[see also][]{kazantsev2001sensitivity}.

Noting that the sensitivity varies widely among short orbits, we then turned to one or a few orbits of long period \citep{lasagna2020sensitivity}. The hypothesis is that a long enough orbit captures all the relevant dynamical events, such as the interactions between scales and the transfer of energy among them \citep{vanveen2019time}, at a resolution sufficient for its probability distribution to model that of the chaotic state. The results in low-dimensional systems were encouraging. In the Lorenz equations and in a minimal Kuramoto--Sivashinsky system, orbits about two orders of magnitude longer than the shortest cycles, converged from near recurrences, have period averages and Floquet exponents that appear to approach, as the period grows, the long-time averages and the Lyapunov exponents of chaotic trajectories. The sensitivities of these orbits also converge as the period grows, but their limit is not necessarily the response of the chaotic state to finite changes of the parameters. Orbits may also meet bifurcations as the parameters change, and at a fold their sensitivities diverge, producing power-law tails in the probability density of the sensitivities. High-dimensional systems may behave better, since their statistics may respond smoothly to the parameters, as in the thermodynamic limit of \citet{ruelle1999smooth}. For turbulent flows the main question remains open, namely whether the statistics of a single long orbit approach those of the flow as the period grows. The yardstick is a turbulent segment of the same duration, and the test is demanding in a flow with rare bursts, since, unlike a library, a single long orbit must visit both the bulk and the tail on its own, in the right proportion. Matching the statistics is also a necessary condition for sensitivities, since an orbit that misses the means gives no reason to trust the derivative of its means. A related question is whether the Floquet exponents of long orbits approach the Lyapunov exponents of the flow, or whether the orbits that a search finds are special ones, perhaps quieter than the flow. More fundamentally, if a long orbit is to stand for the turbulent flow, one would like to know how well it represents turbulent trajectories in their entirety, that is, how close the flow stays to it, and how this changes as the period grows. Answering these questions requires long orbits, but computing them with the standard methods is hard \citep[see the review of][]{ashtari2026computing}.

Shooting has become the dominant method to locate these solutions, largely because it can be built on existing codes for direct numerical simulation (DNS). Matrix-free Newton--Krylov methods, developed for steady states \citep{edwards1994krylov}, with the generalised minimal residual method (GMRES) \citep{saad1986gmres} for the Newton systems and a trust region for the steps, made it practical \citep{viswanath2007recurrent}. Time integration itself acts as a preconditioner for these systems \citep{tuckerman2019order}. Errors in the initial state, however, grow exponentially over the period, so that the radius of convergence of Newton's method shrinks as the period and the instability of the orbit grow \citep{chandler2013invariant, budanur2017relative}. Multiple shooting limits this growth by splitting the period into segments joined by continuity conditions, at the cost of more unknowns \citep{sanchez2010multiple}, and has been used to find longer orbits of pipe flow \citep{budanur2017relative}.

Variational methods \citep{lan2004variational} avoid the exponential growth of errors by treating the whole orbit as a single unknown, a closed loop in state space or, equivalently, a field over space and time, which is deformed until it satisfies the governing equations everywhere along it (figure~\ref{fig:residual}). They are closely related to multiple shooting, since multiple shooting with one time step per segment is itself a discretisation of the whole loop. Which of the two is more robust in practical terms remains a point of discussion in the community. \citet{lan2004variational} deformed the loop by a Newton descent, which converges quadratically near an orbit and has converged very long orbits of low-dimensional systems \citep{lasagna2020sensitivity}, where the whole linear system can be factorised. For a three-dimensional flow, however, the loop is a four-dimensional field, whose Jacobian is far too large to be stored and factorised, and Newton's method on the loop has been judged too expensive for fluid flows \citep{farazmand2016adjoint, azimi2022constructing, zheng2025ghost}. \citet{fazendeiro2010unstable} instead minimised the residual of the loop by gradient descent, which needs no linear solve, and found periodic orbits of weak three-dimensional turbulence, on tens of thousands of processor cores. Adjoint-based methods descend the residual in the same spirit, with a gradient that only requires the application of a differential operator local in time \citep{farazmand2016adjoint, azimi2022constructing, parker2022variational}. They converge from rough guesses, but only linearly \citep{ashtari2023identifying}.

The alternative is to keep Newton's method and solve the linear systems iteratively, but the obstacle is then preconditioning \citep{knoll2004jacobian}. This problem has long been faced in computational fluid dynamics by the harmonic-balance and time-spectral methods, which solve for a whole period of a periodic flow at once \citep{hall2002computation, mcmullen2006demonstration, thomas2013compact, leffell2017time}. Their linear systems couple the instants of the period through the spectral derivative in time, which is diagonal on the harmonics, and the points of the flow at each instant through the spatial operator. The approximate factorisation of \citet{thomas2013compact} replaces the sum of the two operators by their product, so that the two factors can be inverted separately, the temporal one at each point and the spatial one at each instant, and \citet{mundis2017toward} and \citet{ramezanian2020order} use this product as the preconditioner of GMRES. In these applications a few harmonics usually suffice \citep{hall2002computation, mcmullen2006demonstration}. Harmonic balance has more recently reached the fluid mechanics community, to compute limit cycles \citep{douglas2021nonlinear, sierra2022efficient}, their sensitivity \citep{sierra2021adjoint} and the response of transitional flows to forcing periodic in time and in the spanwise direction \citep{rigas2021nonlinear, poulain2024adjoint}, and, in linear form, to analyse the amplification of perturbations about periodic flows \citep{padovan2020analysis}. Its preconditioners there treat each harmonic separately, linearised about the mean flow, and work best when the harmonics interact weakly \citep{rigas2021nonlinear, douglas2021nonlinear}. Our problem, however, is more complex. The unstable periodic orbits of turbulent flows are multiscale in space and time, their harmonics interact strongly, and the number of harmonics they need grows in proportion to the period, so that a long orbit may need hundreds or thousands of them, not just a handful. Making Newton's method on the loop practical for such orbits is the first purpose of this paper.

To this end, we build on this approximate factorisation, which we use as the preconditioner of flexible GMRES within a trust-region Newton method on the loop, and implement the method on graphics processing units (GPUs). The second purpose is to use the method to explore the questions above in two-dimensional Kolmogorov flow at a Reynolds number of 40, a standard testbed for the invariant solutions of turbulence \citep{chandler2013invariant, farazmand2016adjoint, parker2022variational, page2024recurrent}. As in our earlier work, we do not seek a complete hierarchy of orbits, but the properties of the long orbits that searches from near recurrences can find. Sensitivities, the goal that motivates this work, are not computed here. We examine how these searches depend on their initial loops, take a first step towards a comparison with multiple shooting from the same seeds, compare the statistics and the instability of the long orbits with those of the flow, and measure how close the flow stays to them, at this one Reynolds number.

Section~\ref{sec:formulation} formulates the search for orbits as a space-time root-finding problem, and \S\ref{sec:newton} presents the method. Sections~\ref{sec:guesses} and~\ref{sec:performance} analyse the initial loops and the convergence of the searches. Section~\ref{sec:searches} reports the orbits found, \S\ref{sec:shooting} verifies them by multiple shooting and takes a first step towards comparing the two methods, and \S\ref{sec:statistics} relates the long orbits to the turbulent flow. Section~\ref{sec:conclusions} concludes.

\section{Formulation}
\label{sec:formulation}

We introduce Kolmogorov flow, its vorticity formulation and its symmetries (\S\ref{sec:kolmogorov}), and pose the computation of its relative periodic orbits as a root-finding problem on a space of closed loops (\S\ref{sec:spacetime}).

\subsection{Kolmogorov flow}
\label{sec:kolmogorov}

We consider an incompressible fluid in a doubly periodic square of side $L$, driven along $x$ by a body force of amplitude $\chi$ per unit mass that varies sinusoidally in $y$, with $n$ wavelengths across the domain. With lengths scaled by $L/2\pi$ and velocities by $(\chi L/2\pi)^{1/2}$, as in \citet{chandler2013invariant}, the domain becomes the torus $\mathbb{T}^2 = [0, 2\pi)^2$, with position $\bx = (x, y)$, and the Reynolds number is $\Rey = (\chi^{1/2}/\nu)(L/2\pi)^{3/2}$, with $\nu$ the kinematic viscosity. We work directly with the vorticity $\eta = \partial_x v - \partial_y u$ of the velocity $\bu = (u, v)$, for which we write $\eta$ rather than the customary $\omega$, reserved for the frequency $\omega = 2\pi/T$ of an orbit of period $T$. It obeys
\begin{equation}
\partial_t \eta = \cN(\eta) , \qquad
\cN(\eta) = -\bu[\eta]\cdot\nabla\eta + \frac{1}{\Rey}\nabla^2\eta - n\cos(ny) ,
\label{eq:vorticity}
\end{equation}
where $\bu[\eta] = \nabla^{-2}(\nabla^\perp\eta)$ is the velocity of zero mean induced by $\eta$, with $\nabla^{-2}$ the inverse Laplacian on the fields of zero mean, and $\nabla^\perp = (-\partial_y, \partial_x)$ the skew gradient. In this form incompressibility holds by construction, which spares us from dealing with the pressure \citep{ashtari2023identifying}, a problem we leave to future work.

We take $n = 4$ and $\Rey = 40$, a widely studied configuration (\S\ref{sec:introduction}), so that our orbits can be compared with those of other methods. The laminar flow is then unstable and the flow is weakly chaotic, spending most of its time in quiet states in the bulk of the attractor, which it leaves in intermittent bursts of dissipation \citep{platt1991investigation, chandler2013invariant, farazmand2017variational, page2024exact}. Denoting by $[\phi] = (2\pi)^{-2}\int_{\mathbb{T}^2} \phi\,\dd\bx$ the average of a function $\phi$ over the domain, the kinetic energy, the energy input and the dissipation,
\begin{equation}
E = \tfrac12\big[|\bu|^2\big] , \qquad
I = \big[u\sin(ny)\big] , \qquad
D = \frac{1}{\Rey}\big[|\nabla\bu|^2\big] ,
\end{equation}
satisfy $\dd E/\dd t = I - D$, and we report them relative to their laminar values, $E_\mathrm{lam} = \Rey^2/4n^4$ and $I_\mathrm{lam} = D_\mathrm{lam} = \Rey/2n^2$.

The equations are equivariant under the translations $\mathcal{T}_\ell : \eta(x, y) \mapsto \eta(x + \ell, y)$ along the force, which form a continuous group, and under a discrete shift-reflect, whose square is a translation by $2\pi/n$ across the force, and a rotation by $\pi$ \citep[see][]{chandler2013invariant}.

\subsection{Relative periodic orbits as a space-time problem}
\label{sec:orbitsdef}
\label{sec:spacetime}

Because of the continuous symmetry, the recurrent solutions of \eqref{eq:vorticity} are relative periodic orbits, which repeat after a period $T$ up to a translation by a shift $\ell$,
\begin{equation}
\eta(t + T) = \mathcal{T}_\ell\,\eta(t) .
\label{eq:rpo}
\end{equation}
The discrete symmetries permit further orbits, which repeat up to a translation across the force or a shift-reflect, but as in \citet{parker2022variational} and \citet{page2024recurrent} we consider only orbits of the form \eqref{eq:rpo}.

The unknown period $T$ and shift $\ell$, or equivalently the frequency $\omega = 2\pi/T$ and the drift speed $c = -\ell/T$, are built into the equations by a change of variables. In the frame that drifts along $x$ at the speed $c$ the orbit is periodic, and in the rescaled time $s = \omega t$ its period is $2\pi$, whatever $T$. By the chain rule, and since $\cN$ commutes with translations, \eqref{eq:vorticity} becomes
\begin{equation}
\omega\,\partial_s\eta - c\,\partial_x\eta = \cN(\eta) ,
\label{eq:drifting}
\end{equation}
in which $\omega$ and $c$ are unknowns together with the field.

In these variables the orbit is periodic in both directions of space and in the rescaled time, and thus a real field on the three-dimensional torus $\mathbb{T}^3 = [0, 2\pi)^3$, which traces a closed loop in state space. We seek it among the loops of zero spatial mean at every instant, since the vorticity of a periodic velocity field has zero mean,
\begin{equation}
\cU = \Big\{ \eta : \mathbb{T}^3 \to \mathbb{R} \ :\ \big[\eta(\cdot, s)\big] = 0 \Big\} ,
\label{eq:loops}
\end{equation}
where the fields are square integrable together with the derivatives that appear in \eqref{eq:drifting}. We equip $\cU$ with the inner product
\begin{equation}
\langle \eta_1, \eta_2\rangle = \frac{1}{(2\pi)^3}\int_{\mathbb{T}^3} \eta_1(\bx, s)\,\eta_2(\bx, s) \,\dd\bx\,\dd s ,
\label{eq:inner}
\end{equation}
and the norm $\lVert\eta\rVert^2 = \langle\eta, \eta\rangle$. At one instant we use, with the star the complex conjugate, the inner product of two fields $(\eta_1, \eta_2) = (2\pi)^{-2}\int_{\mathbb{T}^2} \eta_1^*(\bx)\,\eta_2(\bx)\,\dd\bx$ and its norm $\lVert \eta(s)\rVert^2 = (\eta(s), \eta(s))$, so that the average of $\lVert\eta(s)\rVert^2$ over the period is $\lVert\eta\rVert^2$. Measuring the vorticity rather than the velocity, both products weight the small scales more heavily than the energy norm would, which suits a residual dominated by temporal and spatial derivatives.

\begin{figure}[t]
\centering
\includegraphics[width=\textwidth]{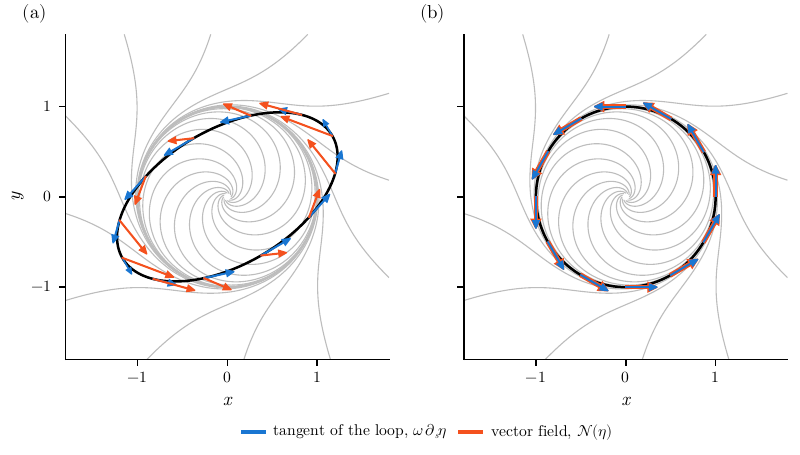}
\caption{The space-time residual for the normal form of a Hopf bifurcation, $\dot x = x - y - x(x^2 + y^2)$, $\dot y = x + y - y(x^2 + y^2)$, whose limit cycle is the unit circle. (a) A loop that is not an orbit, an ellipse with frequency $\omega = 0.8$, along which its tangent (blue) and the vector field (orange) differ by the residual. (b) The orbit, to which the vector field is everywhere tangent. Grey lines are trajectories of the system.}
\label{fig:residual}
\end{figure}

The frequency enters the problem through its logarithm $\gamma = \log\omega$, which keeps it positive during the iterations and makes its changes relative to its value, and the unknowns are collected in
\begin{equation}
p = (\eta, \gamma, c) \in \cP = \cU\times\mathbb{R}\times\mathbb{R} , \qquad
\langle p_1, p_2\rangle = \langle\eta_1, \eta_2\rangle + \gamma_1\gamma_2 + c_1c_2 .
\label{eq:unknowns}
\end{equation}
The orbits are then the zeros of the space-time residual
\begin{equation}
r(p) = \omega\,\partial_s\eta - c\,\partial_x\eta - \cN(\eta) , \qquad \omega = \mathrm{e}^{\gamma} .
\label{eq:residual}
\end{equation}
The residual \eqref{eq:residual} maps $\cP$ into $\cU$, since the derivatives and the equivariant operator $\cN$ preserve the zero mean and the periodicity, and measures how far a candidate loop is from satisfying \eqref{eq:drifting}. Geometrically, it is the difference between the tangent $\omega\,\partial_s\eta$ of the loop in state space and the vector field $\cN(\eta) + c\,\partial_x\eta$ of the drifting frame evaluated on it (figure~\ref{fig:residual}). Similar formulations, in a rescaled time and, for relative solutions, a frame that follows their symmetry, have been used for the periodic orbits of the Kuramoto--Sivashinsky equation \citep{lan2004variational, yang2015numerical}, the relative periodic orbits of the cubic--quintic Ginzburg--Landau equation, whose phase rotates over the period \citep{yang2015numerical}, and the travelling waves and relative periodic orbits of Kolmogorov flow \citep{farazmand2016adjoint, parker2022variational}.

\section{Newton–Krylov method for the space-time problem}
\label{sec:newton}

We formulate Newton's method for the root-finding problem $r(p) = 0$ in the space $\cP$, independently of the discretisation. We then discretise the loops, precondition the linear systems, solve them by flexible GMRES with the steps restricted to a trust region, and describe the implementation on GPUs and the resolution in time.

\subsection{The Newton system}
\label{sec:newtonsystem}

Newton's method corrects an iterate $p_j = (\eta_j, \gamma_j, c_j)$ by $\delta p_j = (\delta\eta_j, \delta\gamma_j, \delta c_j)$, which cancels the residual to first order. The correction changes the whole loop at once, through the space-time field $\delta\eta_j(x, y, s)$. Expanding \eqref{eq:residual} about $p_j$, with $\omega_j = \mathrm{e}^{\gamma_j}$, gives
\begin{equation}
r(p_j + \delta p_j) = r(p_j) + \cJ_{\eta_j}\,\delta\eta_j + \omega_j\,\partial_s\eta_j\,\delta\gamma_j - \partial_x\eta_j\,\delta c_j + O(\lVert\delta p_j\rVert^2) ,
\label{eq:linexp}
\end{equation}
where
\begin{equation}
\cJ_{\eta_j} = \omega_j\,\partial_s - c_j\,\partial_x - \cL_{\eta_j}(s)
\label{eq:Ju}
\end{equation}
is the vorticity equation linearised about the current loop, and
\begin{equation}
\cL_{\eta_j}(s)\,\delta\eta(s) = -\bu[\eta_j(s)]\cdot\nabla\delta\eta(s) - \bu[\delta\eta(s)]\cdot\nabla\eta_j(s) + \frac{1}{\Rey}\nabla^2\delta\eta(s)
\label{eq:linearised}
\end{equation}
is the linearisation of $\cN$ at the instant $s$, which acts on each instant separately.

Setting \eqref{eq:linexp} to zero leaves the correction undetermined, since the resulting system has two more unknowns than equations, $\delta\gamma_j$ and $\delta c_j$, and $\cJ_{\eta_j}$ is singular at an orbit. The translational symmetries in $s$ and in $x$ imply that a shifted solution is also a solution, so that $\partial_s\eta_j$ and $\partial_x\eta_j$ span, generically, the null space of $\cJ_{\eta_j}$ at an orbit. As in shooting methods, two phase conditions,
\begin{equation}
\langle\partial_s\eta_j, \delta\eta_j\rangle = 0 , \qquad \langle\partial_x\eta_j, \delta\eta_j\rangle = 0 ,
\label{eq:phase}
\end{equation}
exclude these shifts from the correction. They add two rows to $\cJ_{\eta_j}$, which fix the components of the correction along the null space. The derivatives of the residual with respect to $\gamma$ and $c$ add two columns, through which a change of the period and of the drift speed removes the part of the residual outside the range of $\cJ_{\eta_j}$. The bordered Newton system
\begin{equation}
\underbrace{\begin{bmatrix}
\cJ_{\eta_j} & \omega_j\,\partial_s\eta_j & -\partial_x\eta_j \\
\langle\partial_s\eta_j, \cdot\,\rangle & 0 & 0 \\
\langle\partial_x\eta_j, \cdot\,\rangle & 0 & 0
\end{bmatrix}}_{\displaystyle\cJ_j}
\underbrace{\begin{bmatrix} \delta\eta_j \\ \delta\gamma_j \\ \delta c_j \end{bmatrix}}_{\displaystyle\delta p_j}
= \underbrace{-\begin{bmatrix} r(p_j) \\ 0 \\ 0 \end{bmatrix}}_{\displaystyle b_j}
\label{eq:newton}
\end{equation}
is thus invertible at a non-degenerate orbit, although $\cJ_{\eta_j}$ is singular there \citep{govaerts2000numerical}. It has the same bordered structure as the Newton descent of \citet{lan2004variational}, with one more row and column for the drift. For the remainder of this section the subscript $j$ is dropped from the Newton system, and kept only on the current iterate, $\eta_j$, $\omega_j$ and $c_j$.

The Newton system \eqref{eq:newton} is far too large to be formed, and we solve it iteratively by GMRES \citep{saad1986gmres}, right-preconditioned by an operator $\cB$ that approximates $\cJ$. GMRES minimises the residual of the system over a Krylov space and needs only products of $\cJ$ with vectors (\S\ref{sec:krylov}).

\subsection{Discretisation in space and time}
\label{sec:discretisation}

The preconditioner $\cB$ is built from the structure of $\cJ$ on the discretised loop, which we therefore describe first. We replace $\cU$ by the space spanned by the Fourier modes with wavenumbers $\bk = (k_x, k_y)$, $|k_x|, |k_y| \le K$, and temporal harmonics $|m| \le M$, in which a loop reads
\begin{equation}
\eta(x, y, s) = \sum_{k_x = -K}^{K}\ \sum_{k_y = -K}^{K}\ \sum_{m = -M}^{M} \tilde\eta_{\bk m}\,\mathrm{e}^{\ii(k_x x + k_y y + ms)} ,
\label{eq:expansion}
\end{equation}
where the coefficients of $\bk = 0$ vanish. We take $K = 33$, and store the loop by the spatial Fourier coefficients of its vorticity at the $N_s = 2M + 1$ equispaced instants $s_l = 2\pi l/N_s$, $l = 0, \ldots, N_s - 1$. Derivatives are applied spectrally, the one in $s$ by fast Fourier transforms in time. The nonlinear term is formed at each instant on a physical grid of $100\times100$ points, with the aliasing removed by the 3/2 rule, a resolution comparable with that of previous work at this Reynolds number \citep{lucas2015recurrent}. In time the products are not dealiased, since dealiasing them leaves the residual practically unchanged, as we verified on the initial loops of \S\ref{sec:guesses}.

\subsection{Preconditioning the Newton systems}
\label{sec:preconditioning}

GMRES converges quickly when the eigenvalues of the operator are clustered away from the origin, and slowly when they fill a wide region of the complex plane \citep[Lecture~35]{trefethen1997numerical}, as those of $\cJ$ do (\S\ref{sec:spectrum}). The preconditioner should thus be close enough to $\cJ$ that the preconditioned operator $\cJ\cB^{-1}$ is close to the identity, yet far easier to invert than $\cJ$. It acts on the field and on the two parameters of the correction separately. On the parameters it scales $\delta\gamma$ and $\delta c$ by the norms of their columns in \eqref{eq:newton}, $s_\gamma = \lVert\omega_j\,\partial_s\eta_j\rVert$ and $s_c = \lVert\partial_x\eta_j\rVert$, so that these columns of the preconditioned operator have unit norm. On the field, its block $\cB_\eta$ follows from the structure of $\cJ_\eta$, the sum of three terms,
\begin{equation}
\cJ_\eta\,\delta\eta =
\underbrace{\omega_j\,\partial_s\delta\eta}_{\displaystyle\Jt\,\delta\eta}
\ \underbrace{-\,c_j\,\partial_x\delta\eta - \frac{1}{\Rey}\nabla^2\delta\eta}_{\displaystyle\Jxo\,\delta\eta}
+ \underbrace{\bu[\eta_j]\cdot\nabla\delta\eta + \bu[\delta\eta]\cdot\nabla\eta_j}_{\displaystyle\Jc\,\delta\eta} ,
\label{eq:split}
\end{equation}
the time derivative, the drift and the viscous term, and the convective term. The time derivative couples all instants but acts on each spatial Fourier mode separately. The drift and the viscous term are diagonal in the spatial Fourier modes and act on each instant separately. The convective term acts on each instant separately but couples all spatial scales and, since its coefficients vary along the period, all temporal harmonics. Which of these couplings is simple depends on the representation of the loop. In the time domain, where the loop is represented by its fields at the instants, as it is stored (\S\ref{sec:discretisation}), the field block reads
\begin{equation}
\cJ_\eta =
\underbrace{\omega_j
\begin{bmatrix}
d_{00} & \cdots & d_{0,N_s-1} \\
\vdots & & \vdots \\
d_{N_s-1,0} & \cdots & d_{N_s-1,N_s-1}
\end{bmatrix}}_{\Jt:\ \text{dense}}
+
\underbrace{\begin{bmatrix}
\Jxo + \Jc(s_0) & & \\
& \ddots & \\
& & \Jxo + \Jc(s_{N_s-1})
\end{bmatrix}}_{\Jxo + \Jc:\ \text{block diagonal}} ,
\label{eq:Jinstants}
\end{equation}
with $d_{ll'}$ the entries of the differentiation matrix of trigonometric interpolation, as in time-spectral methods \citep{mundis2017toward}. In the frequency domain, where the loop is represented by its space-time Fourier modes \eqref{eq:expansion}, it reads instead
\begin{equation}
\tilde\cJ_\eta =
\underbrace{\begin{bmatrix}
-\ii\omega_jM + \tildeJxo & & \\
& \ddots & \\
& & \ii\omega_jM + \tildeJxo
\end{bmatrix}}_{\Jt + \Jxo:\ \text{diagonal}}
+
\underbrace{\begin{bmatrix}
\tilde\cJ_{\mathrm{conv},0} & \cdots & \tilde\cJ_{\mathrm{conv},-2M} \\
\vdots & \ddots & \vdots \\
\tilde\cJ_{\mathrm{conv},2M} & \cdots & \tilde\cJ_{\mathrm{conv},0}
\end{bmatrix}}_{\Jc:\ \text{dense}} ,
\label{eq:Jfrequencies}
\end{equation}
where the block in row $m$ and column $m'$ of the convective term is $\tilde\cJ_{\mathrm{conv},m-m'}$, the harmonic $m - m'$ of the periodic operator $\Jc(s)$. No representation makes the whole field block simple, but each part is simple in one of the two.

In the space-time Fourier modes, the diagonal part $\cJ_0 = \Jt + \Jxo$ suggests the simplest preconditioner, $\cB_0 = \cJ_0$, which multiplies the mode $\mathrm{e}^{\ii(\bk\cdot\bx + ms)}$ by
\begin{equation}
\lambda_{\bk m} = \ii(\omega_j m - c_j\,k_x) + \frac{|\bk|^2}{\Rey} .
\label{eq:lambda}
\end{equation}
These numbers fill a region of the complex plane that extends to $2K^2/\Rey$ along the real axis and to about $\pm\omega_j M$ along the imaginary axis, on which GMRES converges slowly.

This is the space-time form of Stokes preconditioning, in which the inverse of the viscous operator preconditions the Newton systems of steady states and travelling waves, and which weakens as advection strengthens \citep{tuckerman1989steady, tuckerman2019order}. A preconditioner of this kind, which keeps only the highest derivatives in space and time, suffices for the Kuramoto--Sivashinsky and cubic--quintic Ginzburg--Landau equations in one space dimension \citep{yang2015numerical}, as we also found for the Kuramoto--Sivashinsky equation, but in Kolmogorov flow the convective term dominates the spectrum of the Jacobian (\S\ref{sec:spectrum}) and couples the temporal harmonics strongly. In the harmonic-balance computations of weakly forced shear flows, by contrast, the harmonics interact weakly, and a block-diagonal preconditioner, each block linearised about the mean flow, is efficient \citep{rigas2021nonlinear}. Here such a preconditioner is not enough, whether about the mean shear or about the time-averaged flow (not shown), since the advection by the loop changes in time.

A preconditioner for Kolmogorov flow must therefore retain the convective term, as \citet{tuckerman2019order} suggest for Stokes preconditioning, and retain it about the loop itself. Writing $\cJ_\eta = \Jt + \Jx$, with $\Jx = \Jxo + \Jc$ as the spatial part, we follow the approximate factorisation of the time-spectral solvers \citep{thomas2013compact}, used as a preconditioner of GMRES by \citet{mundis2017toward} and \citet{ramezanian2020order}, and replace this sum by a product of a temporal factor, diagonal on the harmonics, and a spatial factor, block diagonal on the instants. Both factors are shifted by a positive number $\sigma$, the inverse of a pseudo-time step in the time-spectral solvers \citep{thomas2013compact}. The shift makes the temporal factor invertible, by moving the zero of the time derivative on the mean harmonic $m = 0$ to $\sigma$, while in practice we found the spatial factor to remain invertible for the values of $\sigma$ we used. Divided by $\sigma$, their product
\begin{equation}
\cB_\eta = \frac{1}{\sigma}\,(\Jt + \sigma\mathsf{I})(\sigma\mathsf{I} + \Jx) = \cJ_\eta + \sigma\mathsf{I} + \frac{1}{\sigma}\Jt\Jx ,
\label{eq:B4}
\end{equation}
contains $\cJ_\eta$, with the error $\sigma\mathsf{I} + \Jt\Jx/\sigma$ of the factorisation. The inverse of $\cB_\eta$ inverts each factor in its own representation,
\begin{equation}
\cB_\eta^{-1} = \sigma
\underbrace{\begin{bmatrix}
\cX_0^{-1} & & \\
& \ddots & \\
& & \cX_{N_s-1}^{-1}
\end{bmatrix}}_{\text{time domain}}
\mathcal{F}_t^{-1}
\underbrace{\begin{bmatrix}
(\sigma - \ii\omega_jM)^{-1} & & \\
& \ddots & \\
& & (\sigma + \ii\omega_jM)^{-1}
\end{bmatrix}}_{\text{frequency domain}}
\mathcal{F}_t ,
\label{eq:B4inverse}
\end{equation}
with $\cX_l = \sigma\mathsf{I} + \Jx(s_l)$ and $\mathcal{F}_t$ the discrete Fourier transform in time. Read from right to left, $\cB_\eta^{-1}$ applies the Fourier transform in time, divides each harmonic $m$ by $\sigma + \ii\omega_jm$, which inverts the temporal factor, transforms back to the instants, and then solves the spatial block $\cX_l$ of each instant independently, which inverts the spatial factor, convection included. The shift decides which modes are well preconditioned. We tested fixed shifts between 0.5 and 4 along whole searches, and found that $\sigma = 2.5$ suits the short loops and a smaller shift the longer ones (not shown).

It remains to invert the spatial blocks, one per instant,
\begin{equation}
\cX_l\,\psi_l = \big(\sigma\mathsf{I} + \Jxo + \Jc(s_l)\big)\,\psi_l = g_l , \qquad l = 0, \ldots, N_s - 1 ,
\label{eq:blocks}
\end{equation}
which we do by ten steps of GMRES, right-preconditioned by $\cB_{\partial x} = \sigma\mathsf{I} + \Jxo$. This operator multiplies each spatial Fourier mode $\mathrm{e}^{\ii\bk\cdot\bx}$ by $\zeta_{\bk} = \sigma - \ii c_jk_x + |\bk|^2/\Rey$ and is thus inverted exactly. With $\psi_l$ and $g_l$ represented by their spatial Fourier modes, GMRES sees the operator
\begin{equation}
\begin{aligned}
\cX_l\,\cB_{\partial x}^{-1} &= \big(\sigma\mathsf{I} + \Jxo + \Jc(s_l)\big)\,(\sigma\mathsf{I} + \Jxo)^{-1} = \mathsf{I} + \Jc(s_l)\,(\sigma\mathsf{I} + \Jxo)^{-1} \\
&= \mathsf{I} +
\underbrace{\Big(\mathcal{F}_x
\begin{bmatrix} \bu[\eta_j(s_l)]\cdot(\cdot) & (\cdot)\cdot\nabla\eta_j(s_l) \end{bmatrix}
\mathcal{F}_x^{-1}
\begin{bmatrix} \nabla \\ \bu[\,\cdot\,] \end{bmatrix}\Big)}_{\displaystyle\Jc(s_l)}
\underbrace{\begin{bmatrix}
\ddots & & \\
& \zeta_{\bk}^{-1} & \\
& & \ddots
\end{bmatrix}}_{\displaystyle(\sigma\mathsf{I} + \Jxo)^{-1}} ,
\end{aligned}
\label{eq:blockprec}
\end{equation}
with $\mathcal{F}_x$ the discrete Fourier transform in space. Read from right to left, its second term divides each spatial Fourier mode by $\zeta_{\bk}$, forms the gradient and the induced velocity of the result, transforms them to the physical grid, where they multiply the velocity and the vorticity gradient of the loop at the instant $s_l$, and transforms the products back. Since convection involves at most first derivatives while $\zeta_{\bk}$ grows as $|\bk|^2/\Rey$, the perturbation of the identity decays at high wavenumbers, and GMRES reduces the residual of every preconditioned block by about a decade in ten steps, at every instant and at the spatial resolutions we tested, which has proved sufficient. Including also the advection by the mean shear in this preconditioner slowed the inner solves (not shown). The $N_s$ block systems are independent but solved together. Every inner step applies all the blocks at once, as one application of the linearised operator to the whole loop, so that the inner solve takes ten such applications.

The borders of the Newton system are not approximated. The whole preconditioner, $\cB = \operatorname{diag}(\cB_\eta, s_\gamma, s_c)$, combines the field preconditioner with the scaling of the parameters, and the preconditioned Newton operator reads
\begin{equation}
\cJ\cB^{-1}
=
\begin{bmatrix} \cJ_\eta\cB_\eta^{-1} & 0 & 0 \\ 0 & 1 & 0 \\ 0 & 0 & 1 \end{bmatrix}
+
\underbrace{\begin{bmatrix}
0 & \omega_j\,\partial_s\eta_j/s_\gamma & -\partial_x\eta_j/s_c \\
\langle\partial_s\eta_j, \cB_\eta^{-1}\,\cdot\,\rangle & -1 & 0 \\
\langle\partial_x\eta_j, \cB_\eta^{-1}\,\cdot\,\rangle & 0 & -1
\end{bmatrix}}_{\displaystyle\mathcal{E}} ,
\label{eq:bordered}
\end{equation}
the preconditioned field block, extended by the identity on the parameters, plus a matrix $\mathcal{E}$ that is nonzero only in its last two columns and rows and has rank at most four. In practice GMRES converges on the Newton system about as it would on the field block alone, with a few more steps (not shown), and the preconditioner needs to approximate only $\cJ_\eta$.

The preconditioner is necessary for convergence (\S\ref{sec:spectrum}), but it is expensive. With ten inner steps, an application of $\cB^{-1}$ takes as long as 13 to 21 products with $\cJ$ on an NVIDIA H200 GPU, as detailed in appendix~\ref{app:speedup}, so that the inner steps take most of the time spent on the operators in an Arnoldi step of GMRES (\S\ref{sec:krylov}). This time is the price of retaining convection.

\subsection{Krylov solution and the hookstep}
\label{sec:krylov}

The preconditioner $\cB$ enters on the right of $\cJ$, as in \eqref{eq:bordered}, and the correction is written as
\begin{equation}
\delta p = \cB^{-1}\delta z ,
\label{eq:changeofvariables}
\end{equation}
and GMRES solves
\begin{equation}
\delta z_k = \operatorname*{arg\,min}_{\delta z\,\in\,\mathcal{K}_k} \lVert b - \cJ\cB^{-1}\delta z\rVert , \qquad
\mathcal{K}_k = \operatorname{span}\{b, \cJ\cB^{-1}b, \ldots, (\cJ\cB^{-1})^{k-1}b\} .
\label{eq:gmresprec}
\end{equation}
Since $\cJ\cB^{-1}\delta z = \cJ\,\delta p$, right preconditioning leaves unchanged the residual of the Newton system that GMRES minimises, and changes only the space in which the correction is sought. Each Arnoldi step adds to the basis of $\mathcal{K}_k$ a loop, a space-time field together with two scalars, at the cost of one application of $\cB^{-1}$ and one product with $\cJ$, and orthogonalises it against the previous ones by two passes of the Gram--Schmidt process. The orthonormal basis $Q_k$ satisfies the Arnoldi relation $\cJ\cB^{-1}Q_k = Q_{k+1}H_k$, with $H_k$ upper Hessenberg of size $(k + 1)\times k$, so that writing $\delta z = Q_k y$ turns \eqref{eq:gmresprec} into a small least-squares problem with $H_k$ for the $k$ coordinates $y\in\mathbb{R}^k$, with $\lVert y\rVert = \lVert\delta z\rVert$.

The inner GMRES steps of \S\ref{sec:preconditioning}, however, do not apply $\cB^{-1}$ as a fixed linear operator, so that the preconditioner changes slightly from one Arnoldi step to the next. Standard GMRES tolerates this change only if the inner solutions are accurate, and the searches then often fail with fewer than twenty inner steps. The flexible GMRES of \citet{saad1993flexible} admits a changing preconditioner, and the searches converge with ten inner steps, in about half the wall time of standard GMRES with twenty (not shown). It stores the preconditioned vectors $w_i = \cB_i^{-1}q_i$ of the Arnoldi vectors $q_i$ in a matrix $W_k$, with $\cJ W_k = Q_{k+1}H_k$, so that the correction $\delta p = W_k y$ follows from the same coordinates $y$.

Newton's method converges only from loops close to an orbit. Following \citet{viswanath2007recurrent}, we therefore restrict the correction to a trust region of radius $\Delta$. The change of variables \eqref{eq:changeofvariables}, which leaves the residual of the Newton system unchanged, sets instead the metric in which the length of the correction is measured, $\lVert\delta z\rVert = \lVert\cB\,\delta p\rVert$, similar in spirit to the scaling of the adjoint descent of \citet{farazmand2016adjoint}. With $\cB_0$, for instance, the trust region is an ellipsoid with semi-axes $\Delta/|\lambda_{\bk m}|$, shorter along the high frequencies and wavenumbers that $\cJ$ amplifies. In the coordinates $y$, where $\lVert y\rVert = \lVert\cB\,\delta p\rVert$ for a fixed preconditioner, the step solves the comparatively small hookstep problem
\begin{equation}
\min_{y\in\mathbb{R}^k} \lVert b - \cJ W_k y\rVert \quad\text{subject to}\quad \lVert y\rVert \le \Delta ,
\label{eq:hookstep}
\end{equation}
computed in negligible time through the singular value decomposition of $H_k$.

We measure the progress of a search by the global residual $R(p) = \tfrac12\lVert r(p)\rVert^2$, with $r$ the space-time residual \eqref{eq:residual}, which vanishes only at the orbits and which near $p_j$ is approximated by the linear model $\tfrac12\lVert b - \cJ\,\delta p\rVert^2$. The ratio $\rho$ of the actual reduction of $R$ to the reduction predicted by the model, which follows from \eqref{eq:hookstep} without products with the Jacobian, measures how well the model holds. The radius adapts to $\rho$ by the classical rules of \citet{dennis1983numerical}, and a step is accepted when $\rho > 10^{-3}$. A rejected step is solved again at a smaller radius in the same Krylov space.

The Arnoldi iteration stops when the GMRES relative residual $\lVert b - \cJ\,\delta p\rVert/\lVert b\rVert$ falls below $10^{-2}$, since far from the orbit the Newton system need not be solved exactly. In the campaigns of \S\ref{sec:searches}, with ten to twenty inner steps, a search stops when $\lVert r\rVert$ falls below $10^{-12}$ in double precision or $10^{-5}$ in single precision, after 50 Newton iterations, or when the residual has not halved over the last ten to fifteen iterations.

\subsection{Implementation}
\label{sec:implementation}

The method is implemented in the Julia language. Its solver needs only the residual, the product of the Jacobian with a vector, the application of a preconditioner and an inner product, and so applies to other systems as well. For Kolmogorov flow the product with $\cJ$ applies the linearised equation \eqref{eq:linearised} exactly, while for systems whose linearisation is not available it could be approximated by a finite difference of residuals, as in Jacobian-free Newton--Krylov methods \citep{knoll2004jacobian}. Every operator of Kolmogorov flow is a short sequence of fast Fourier transforms and element-wise operations on one three-dimensional array, which holds the spatial Fourier modes of the vorticity at all the instants. The same code runs on central processing units (CPUs) and on GPUs, but this structure suits GPUs, which reach their throughput only when they have enough independent work to hide the latency of their memory. Time stepping offers such work within each instant and, in multiple shooting, across the segments, but the instants of a segment are computed one after the other. The space-time formulation adds time as a further direction of parallelism, as parallel-in-time methods do. On a GPU the operators run tens to hundreds of times faster than on one CPU core (appendix~\ref{app:speedup}). An Arnoldi step of the searches takes about 0.2~s on an NVIDIA H200 GPU for a loop of a thousand instants.

Without the mean mode, each instant carries $(2K + 1)^2 - 1 = 4\,488$ real unknowns, so that the size of a loop grows in proportion to its period. The longest orbit we found, of period 190.79, was converged on 1\,575 instants, about eight per unit time, with $7.1\times10^6$ unknowns that occupy 28~MB in single precision. The memory that a search needs is set instead by its Krylov space, since flexible GMRES stores two loops for every Krylov vector, and with the present preconditioner loops of period 100 use about a thousand vectors (\S\ref{sec:cost}). The preconditioner reduces the number of Arnoldi steps, and with it the size of the Krylov space and the time of its orthogonalisation, which grows as the square of the number of steps. Most searches ran in single precision, which halves the memory, and on four to eight instants per unit time (\S\ref{sec:protocol}).

\begin{figure}[t]
\centering
\includegraphics[width=\textwidth]{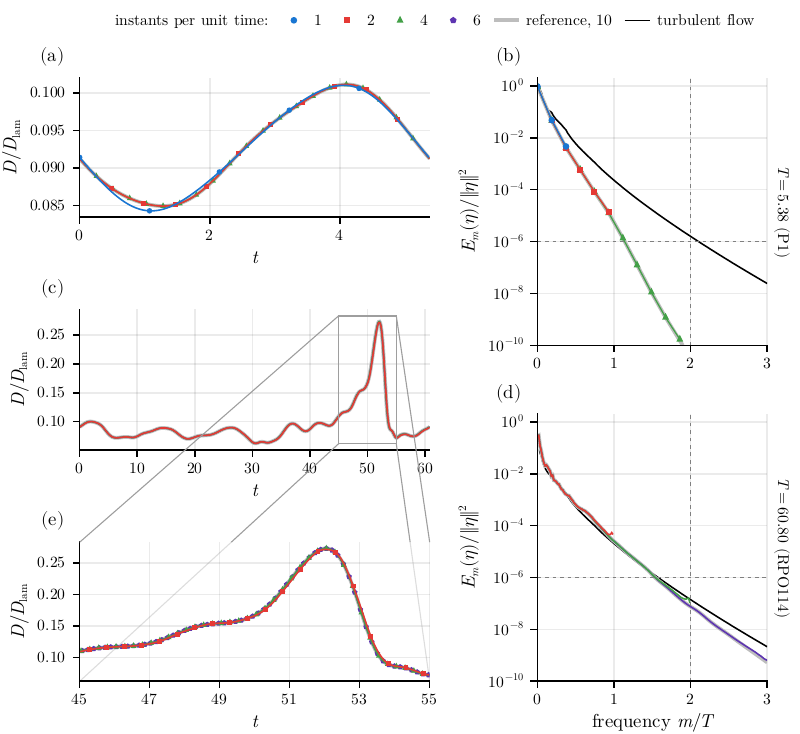}
\caption{The effect of the resolution in time, on the periodic orbit P1 of \citet{chandler2013invariant}, of period 5.38 (a, b), and on RPO114, of period 60.80, which visits a burst (c--e). P1 is converged on one, two and four instants per unit time (blue, red and green) and RPO114 on two, four and six (red, green and purple), both also on about ten (grey). (a, c) The dissipation along the period, and (e) the burst of RPO114 in the window marked in (c). (b, d) The temporal spectrum \eqref{eq:energy}, relative to $\lVert\eta\rVert^2$, over that of the turbulent flow (black), with the level $10^{-6}$ and the highest frequency resolved on four instants per unit time (dashed).}
\label{fig:temporal}
\end{figure}

The number of instants per unit time sets the temporal resolution, which we judge by the temporal spectrum of the loop, the energies
\begin{equation}
E_0(\eta) = \sum_{\bk}|\tilde\eta_{\bk 0}|^2 , \qquad
E_m(\eta) = \sum_{\bk}\big(|\tilde\eta_{\bk m}|^2 + |\tilde\eta_{\bk, -m}|^2\big) , \quad m = 1, \ldots, M ,
\label{eq:energy}
\end{equation}
of its temporal harmonics, which add up to $\lVert\eta\rVert^2$. The temporal spectra of the orbits we converged fall steeply, even for those with the most active dynamics, and on four instants per unit time they span at least six decades. Figure~\ref{fig:temporal} and table~\ref{tab:temporal} illustrate this on P1, a short periodic orbit, converged again on one, two and four instants per unit time, and on RPO114, a relative periodic orbit that visits a burst, converged again on two, four and six instants per unit time. Here and below we name the solutions of \citet{chandler2013invariant} as they did, and the others by their labels in table~S1 of the supplementary material. RPO114 has nearly the temporal spectrum of the turbulent flow, while that of P1 falls much faster. Yet on four instants per unit time, that is 121 temporal harmonics, RPO114 agrees with its version on about ten instants per unit time to about five digits (table~\ref{tab:temporal}), although its dissipation nearly triples during the burst and falls back within two time units (figure~\ref{fig:temporal}e). On six instants per unit time the agreement improves to five to eight digits, and on two it falls to one to three.

\begin{table}
\begin{center}
\def~{\hphantom{0}}
\begin{tabular}{lrrrcccc}
 & $N_s$ & $M$ & $N_s/T$ & $T$ & $\ell$ & $\overline{D}/D_\mathrm{lam}$ & $\max D/D_\mathrm{lam}$ \\[3pt]
P1 & 5 & 2 & 0.9 & \underline{5.3}6885181 & 0 & \underline{0.092}70589 & \underline{0.101}0344 \\
 & 11 & 5 & 2.0 & \underline{5.380}44894 & 0 & \underline{0.092804}88 & \underline{0.101152}1 \\
 & 21 & 10 & 3.9 & \underline{5.38039990} & 0 & \underline{0.09280455} & \underline{0.1011520} \\
 & 63 & 31 & 11.7 & 5.38039990 & 0 & 0.09280455 & 0.1011520 \\[3pt]
RPO114 & 121 & 60 & 2.0 & \underline{60.8}6203797 & \underline{2.}89346129 & \underline{0.094}07992 & \underline{0.27}15097 \\
 & 243 & 121 & 4.0 & \underline{60.803}69687 & \underline{2.903}49719 & \underline{0.094170}33 & \underline{0.27305}66 \\
 & 375 & 187 & 6.2 & \underline{60.803977}46 & \underline{2.903500}94 & \underline{0.09417014} & \underline{0.27305}01 \\
 & 625 & 312 & 10.3 & 60.80397757 & 2.90350061 & 0.09417014 & 0.2730574 \\

\end{tabular}
\caption{The orbits of figure~\ref{fig:temporal} on $N_s$ instants, $M$ temporal harmonics and $N_s/T$ instants per unit time, with their period, shift, and mean and largest dissipation. Underlined digits agree with the orbits of the first campaign (\S\ref{sec:protocol}), on 63 and 625 instants.}
\label{tab:temporal}
\end{center}
\end{table}

\section{Initial loops from near recurrences}
\label{sec:guesses}

\subsection{Near recurrences of long period}
\label{sec:recurrences}

\begin{figure}[!tp]
\centering
\includegraphics[width=\textwidth]{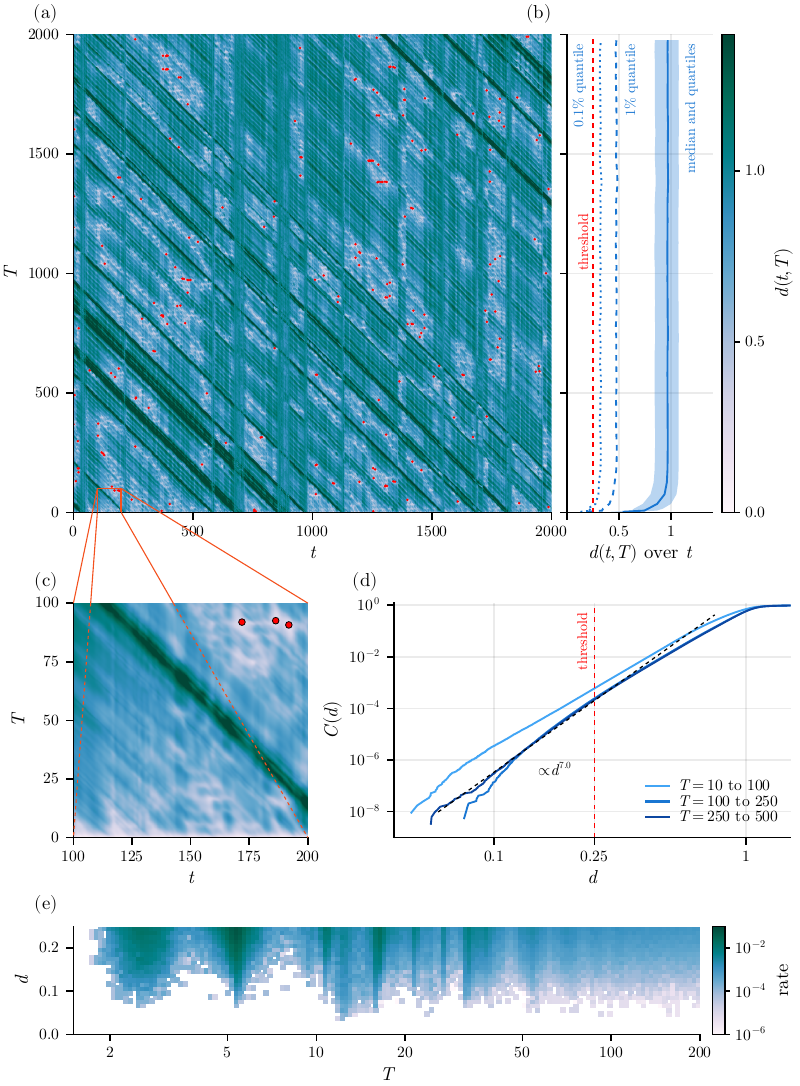}
\caption{Near recurrences of the turbulent flow. (a) The distance \eqref{eq:distance} for $t$ and $T$ up to 2\,000 along one simulation, with its local minima below the threshold 0.25 (red), and (c) the box marked in (a). (b) The median, the quartiles and the 1\% and 0.1\% quantiles of $d(t, T)$ over $t$, in bins of $T$, over sixty-four simulations. (d) The correlation integral $C(d)$, the fraction of pairs of states closer than $d$, in three bands of $T$, with a power law fitted to the pairs with $100 \le T \le 500$ for $0.1 \le d \le 0.4$ (dashed). (e) The rate of near recurrences per unit time, unit period and unit distance. Red dashed lines mark the threshold.}
\label{fig:recurrences}
\end{figure}

As in previous studies \citep{kawahara2001periodic, viswanath2007recurrent, chandler2013invariant, lucas2015recurrent}, we take initial loops from near recurrences of long simulations of \eqref{eq:vorticity}, the local minima of the relative distance
\begin{equation}
d(t, T) = \min_{\ell}\frac{\lVert\mathcal{T}_\ell\,\eta(t) - \eta(t + T)\rVert}{\lVert\eta(t)\rVert} ,
\label{eq:distance}
\end{equation}
which figure~\ref{fig:recurrences}(a) shows along one simulation for $t$ and $T$ up to 2\,000, and figure~\ref{fig:recurrences}(c) in a small window. The distance is homogeneous over this map, in $t$ and, beyond about a hundred time units, in $T$. Two states a long time apart are then as far apart as independent states of the attractor, and the distribution of $d$ over $t$ no longer changes (figure~\ref{fig:recurrences}b). This suggests that recurrences as good as those of moderate period can be found at much longer ones. We keep the minima of $d$ below 0.25 (figure~\ref{fig:recurrences}c), a distance that independent states rarely reach. Using this criterion up to a period of 200, we obtained a database of near recurrences, of which the searches of \S\ref{sec:searches} used $3\,530$ as seeds (\S\ref{sec:protocol}).

Near recurrences concentrate at short periods (figure~\ref{fig:recurrences}e). These close recurrences cluster around the period 5.38 of the periodic orbit P1 of \citet{chandler2013invariant} and its multiples, an orbit that \citet{page2024recurrent} also recover frequently. There are also clusters around the period 12.21 of the relative periodic orbit R19 and, at the shortest periods, just below the period 2.83 of the periodic orbit P2. These represent short orbits that the flow visits often. Above $T\approx40$ these clusters fade, and beyond about a hundred time units the rate of near recurrences per unit time, per unit period and per unit distance becomes nearly invariant with $T$, since the probability that two states come within the threshold, which sets this rate, no longer depends on $T$ (figure~\ref{fig:recurrences}b, d).

The fraction $C(d)$ of pairs closer than $d$ in figure~\ref{fig:recurrences}(d), the probability that two independent states lie within $d$ of each other, is the correlation integral of \citet{grassberger1983characterization}. Its slope, about seven (dashed line) and rising slowly as $d$ falls, estimates the correlation dimension $D_2$ of the attractor without the translation along $x$, below its Kaplan--Yorke dimension of about 10.5, also without the translation, given by the Lyapunov exponents of \S\ref{sec:floquet}. The threshold 0.25 that we chose thus corresponds to a probability of about $2\times10^{-4}$ for independent states.

That this probability does not depend on $T$ may seem somewhat counterintuitive, since in a uniformly hyperbolic flow a segment that nearly closes follows a periodic orbit over its whole length. While the probability that a trajectory stays near a given orbit decays exponentially with the period, as $\mathrm{e}^{-\bar\sigma T}$, with $\bar\sigma$ the real part of the sum of its unstable Floquet exponents (\S\ref{sec:floquet}), the number of orbits also grows exponentially with the period, as $\mathrm{e}^{hT}$, with $h$ the topological entropy \citep[\S 21.3]{chaosbook}, and the two factors balance. At $T = 200$, for instance, taking for both rates the sum of the positive Lyapunov exponents, about 0.27 (\S\ref{sec:floquet}), a trajectory stays near a given orbit with a probability of order $10^{-23}$, but there should be of the order of $10^{23}$ orbits. Kolmogorov flow is not uniformly hyperbolic, however, and this argument is only a guide. What becomes rare with the period is not a near recurrence but a second near recurrence of the same orbit.

\begin{figure}[t]
\centering
\includegraphics[width=\textwidth]{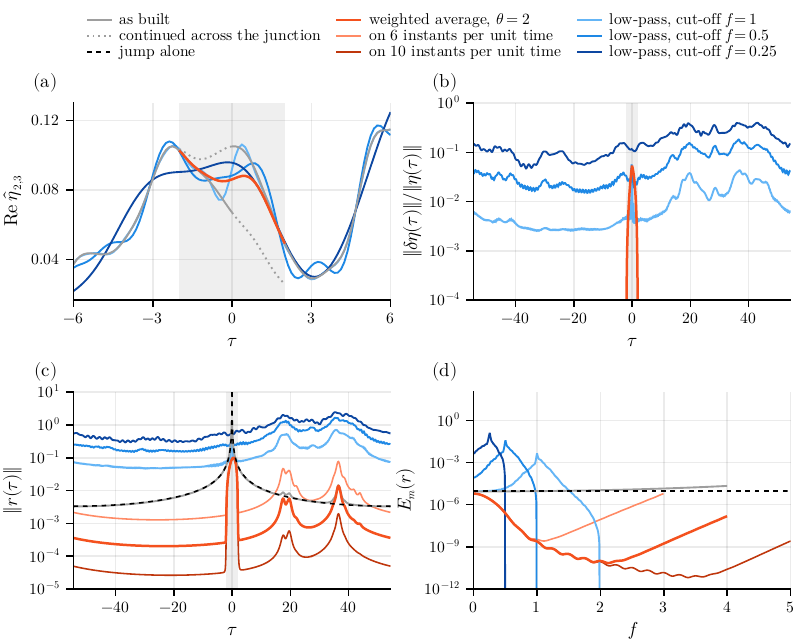}
\caption{Smoothing the jump of an initial loop of period 108.8 that passes through a burst, on eight instants per unit time, by low-pass filters at three cut-offs and by the weighted average \eqref{eq:filter}, against the time $\tau$ from the junction, with the window $|\tau| \le \theta$ shaded. (a) The real part of the vorticity coefficient that jumps most at the junction, $\hat\eta_{2,3}$, of wavenumber $(k_x, k_y) = (2, 3)$. (b) The relative change of the loop at each instant. (c) The norm of the residual at each instant and (d) its temporal spectrum \eqref{eq:energy} against the frequency $f = m/T$, both with the weighted average also on six and ten instants per unit time and with the jump alone (dashed).}
\label{fig:lowpass}
\end{figure}

\subsection{Smoothing the jump of the loop}
\label{sec:smoothing}

A question we asked from the outset is how much convergence depends on the initial loop, and on its residual in particular. The loop given by a near recurrence has a jump of norm $d\,\lVert\eta(t)\rVert$ where its end meets its start. Spectral differentiation in time turns the jump into a peak of the residual at the junction, which decays only as $1/|\tau|$ with the time $\tau$ from the junction (figure~\ref{fig:lowpass}c). The jump makes the loop differ from a smooth one by a sawtooth, whose temporal harmonics have amplitudes $d\,\lVert\eta(t)\rVert/(2\pi m)$. Differentiation in time multiplies harmonic $m$ by $2\pi m/T$, and the jump thus pollutes the whole temporal spectrum \eqref{eq:energy} of the residual, with about the same energy $2d^2\lVert\eta(t)\rVert^2/T^2$ in every harmonic (figure~\ref{fig:lowpass}d).

A common approach is to low-pass filter the loop in time \citep{lan2004variational, azimi2022constructing}. This smooths the jump but also moves the loop off the trajectory at every instant, the more so the lower the cut-off frequency (figure~\ref{fig:lowpass}a, b). It thus creates a residual away from the junction, where that of the loop as built is small (figure~\ref{fig:lowpass}c). Expecting a smaller initial residual to help convergence, we initially adopted instead, and ran most of our searches with, a smoothing localised in time, which replaces the loop near the junction by a weighted average of the trajectories that enter and leave it. It assumes that, far enough from the junction, the loop lies close to an orbit and needs no change. Within a window $|\tau| \le \theta$ about the junction the smoothed loop is
\begin{equation}
\eta^{\theta}(\tau) = \big(1 - \beta(\tau)\big)\,\eta(t + T + \tau) + \beta(\tau)\,\eta(t + \tau) , \qquad
\beta(\tau) = \sin^2\!\Big(\frac{\pi(\tau + \theta)}{4\theta}\Big) , \qquad |\tau| \le \theta .
\label{eq:filter}
\end{equation}
The weight $\beta$ ensures that the loop passes smoothly from the end of the segment to its start, with a continuous derivative in time (figure~\ref{fig:lowpass}a). The weighted average confines the residual to the window, where it leaves a plateau with steep edges (figure~\ref{fig:lowpass}c), and replaces the flat spectrum of the jump by one that falls by several decades (figure~\ref{fig:lowpass}d). Outside the window only the error of the spectral derivative of the sampled trajectory remains. The instants cannot resolve the fastest variations of the trajectory, which end up in the highest harmonics, where the derivative amplifies them most, so that the spectrum rises again near the last harmonics. As the number of instants per unit time grows, this error decreases and the rise moves to higher frequencies and lower energies (figure~\ref{fig:lowpass}c, d). On the initial loops we tested, the weighted average lowers the norm \eqref{eq:inner} of the residual several-fold with respect to the loop as built for half-widths between 1 and 4, while on wider windows the two trajectories separate and the residual grows steeply (not shown). We therefore applied the weighted average with $\theta = 2$ to all the initial loops of the campaigns (\S\ref{sec:protocol}).

\begin{table}
\begin{center}
\def~{\hphantom{0}}
\begin{tabular}{lccccc}
 & & \multicolumn{4}{c}{medians over the searches} \\
 & seeds converged & initial $\lVert r\rVert$ & Newton iterations & hooksteps & total Arnoldi steps \\[3pt]
weighted average & 112 (6) & 0.042 & 16 & 14 & 6\,517 \\
low-pass filter  & 107 (2) & 0.106 & 15 & 13 & 6\,383 \\
as built         & 105 (1) & 0.235 & 16 & 14 & 6\,371 \\

\end{tabular}
\caption{The effect of smoothing on the searches from 224 seeds of the third campaign, half of them the seeds of the 112 distinct orbits it had found by then and half seeds whose first search failed, from loops smoothed by the weighted average \eqref{eq:filter}, low-pass filtered at one cycle per unit time, or left as built. In brackets, the seeds that converge with that smoothing only. The medians are over the 87 seeds that converge to the same orbit with all three, and the hooksteps are the steps on the boundary of the trust region.}
\label{tab:lowpass}
\end{center}
\end{table}

Whether this smoothing matters for convergence is a question we returned to a posteriori. We searched again the seeds of the 112 distinct orbits that the third campaign had found, and as many random seeds of the same periods whose first search had failed, from loops low-pass filtered at one cycle per unit time and from the loops without any smoothing, with otherwise the same settings. The weighted average has only a marginal effect on convergence, compared with the low-pass filter and even with the loops as built (table~\ref{tab:lowpass}). With it the searches converge just a few more seeds than with either, and where they converge they mostly reach the same orbit. Despite initial residuals several times larger, the low-pass filtered loops and the loops as built reach the same orbits in about as many Newton iterations and Arnoldi steps. The residual of the initial loop is therefore not, by itself, a criterion for whether a search converges.

\section{Convergence of the searches}
\label{sec:performance}

We illustrate the behaviour of the method on single searches. The Newton system of one initial loop shows the action of the preconditioner (\S\ref{sec:spectrum}), two searches show how hooksteps and inexact Newton steps share the work and how a search stalls, and a third how the Newton correction differs from the gradient (\S\ref{sec:course}), and the searches of the campaigns give the wall time and the memory (\S\ref{sec:cost}).

\begin{figure}[t]
\centering
\includegraphics[width=\textwidth]{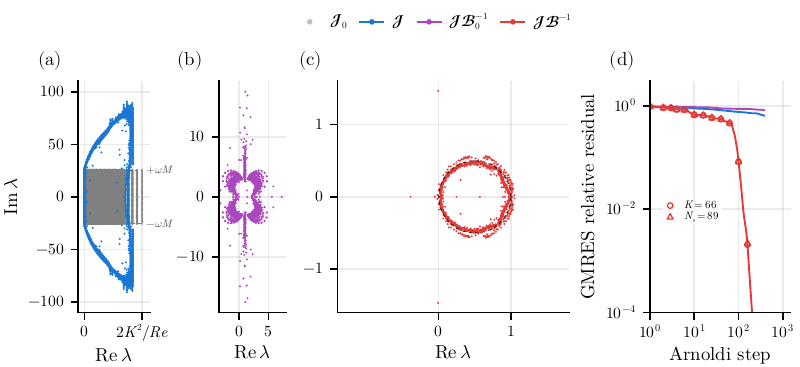}
\caption{The effect of the preconditioner on the Newton system of the initial loop of period 5.4 that converges to P1. (a) The Ritz values of $\cJ$ after 2\,000 Arnoldi steps, over the eigenvalues \eqref{eq:lambda} of $\cJ_0$ (grey), which fill the rectangle of width $2K^2/\Rey$ and half-height $\omega M$. (b) The same for $\cJ\cB_0^{-1}$ and (c) for $\cJ\cB^{-1}$ with $\sigma = 2.5$, with the circle of diameter one centred at one half (dashed). (d) The GMRES relative residual against the Arnoldi step, also with twice the Fourier modes in space (circles) and in time (triangles).}
\label{fig:spectrum}
\end{figure}

\subsection{The Newton systems}
\label{sec:spectrum}

The preconditioner is meant to cluster the eigenvalues away from the origin (\S\ref{sec:preconditioning}). Figure~\ref{fig:spectrum} shows its effect on the Newton system of the initial loop of period 5.4 that converges to P1. The grey dots are the eigenvalues \eqref{eq:lambda} of the diagonal part $\cJ_0$ of the Jacobian, one ladder for each spatial wavenumber with rungs spaced by $\omega$, which together fill a rectangle of width $2K^2/\Rey$, set by the viscous term, and half-height $\omega M$, set by the time derivative (figure~\ref{fig:spectrum}a). Advection by the loop stretches the spectrum of the full Jacobian much further, to imaginary parts far larger than $\omega M$, a sign that advection is fundamental to this problem. The outer part of the spectrum, which the Ritz values approximate, no longer changes between 400 and 2\,000 Arnoldi steps, for the Jacobian as for the two preconditioned operators (not shown). The preconditioner $\cB_0 = \cJ_0$, which inverts the time derivative, the drift and the viscous term but not convection, spreads the spectrum around the origin (figure~\ref{fig:spectrum}b), and GMRES does worse with it than without (figure~\ref{fig:spectrum}d). With the factorised preconditioner the Ritz values gather about the circle of diameter one centred at one half (figure~\ref{fig:spectrum}c), which encloses the eigenvalues of $\cJ_0$ preconditioned in the same way, and GMRES converges within a few hundred Arnoldi steps, a number that does not change when the loop is refined twofold in space or in time (figure~\ref{fig:spectrum}d). The Ritz values of longer loops have the same structure, only more numerous (not shown), and on these loops GMRES first stays on a plateau that lengthens with the period (\S\ref{sec:cost}). On the initial loop of RPO178, the longest orbit that we found, the GMRES relative residual does not change either when the loop is refined by about 50\% in space or in time (not shown).

\subsection{The course of a search}
\label{sec:course}

\begin{figure}[t]
\centering
\includegraphics[width=\textwidth]{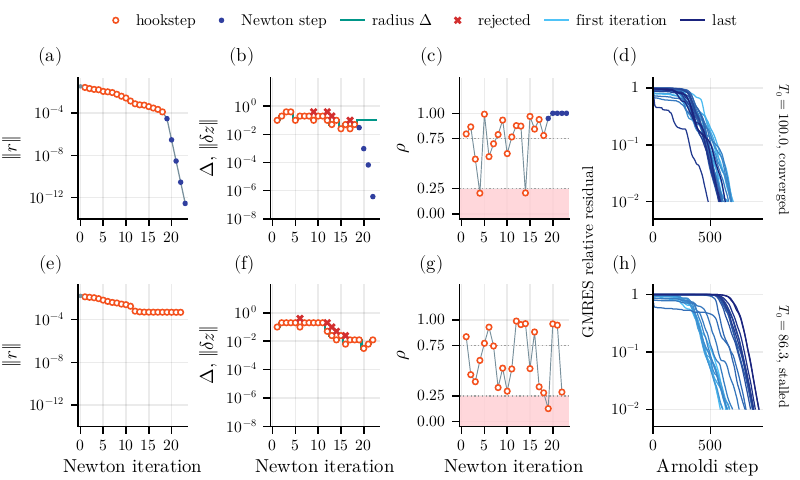}
\caption{The progress of two searches, one per row, from a loop of period 100.0 that converges to RPO168 (a--d) and from one of period 86.3 that stalls (e--h). (a, e) The residual at each Newton iteration, open for hooksteps on the boundary of the trust region and filled for Newton steps. (b, f) The radius $\Delta$ of the trust region and the length of the step, with a cross at each rejected step. (c, g) The ratio $\rho$ of \S\ref{sec:krylov}, with the thresholds 1/4, below which (shaded) the radius shrinks, and 3/4. (d, h) The GMRES relative residual against the Arnoldi step, one line per Newton iteration, light to dark.}
\label{fig:traces}
\end{figure}

Figure~\ref{fig:traces} shows two searches from loops of similar period, one that converges to RPO168 (figure~\ref{fig:traces}a--d) and one that stalls (figure~\ref{fig:traces}e--h). Their course follows from the form of the hookstep. Without preconditioner, a step on the boundary of the trust region is
\begin{equation}
\delta p(\mu) = \big(\cJ^\dagger\cJ + \mu\mathsf{I}\big)^{-1}\cJ^\dagger b ,
\label{eq:hookcurve}
\end{equation}
with $\cJ^\dagger$ the adjoint of $\cJ$ in the inner product \eqref{eq:unknowns} and $\mu > 0$ such that $\lVert\delta p(\mu)\rVert = \Delta$. It tends to the Newton step $\cJ^{-1}b$ on a large trust region and to the direction of steepest descent $\cJ^\dagger b = -\nabla R$ on a small one \citep{dennis1983numerical}, and with the preconditioner the same holds within the Krylov space, in the metric of the preconditioner. The first search thus passes from a slow descent, with steps on the boundary of the trust region, to an inexact Newton iteration whose residual falls by about two decades per iteration, the relative tolerance of GMRES.

Far from an orbit the hookstep thus descends $R$, as the adjoint-based variational methods do \citep{farazmand2016adjoint, parker2022variational}, but on a quadratic model of $R$, so that its steps also move in part along the Newton step, the more so as the model improves. With the robustness of descent thus built in, we did not find a separate stage of descent necessary. Like any descent, the hookstep may converge to minima of $R$ that are not orbits \citep{dennis1983numerical}, as the second search seems to do, never leaving the first phase while its residual stops falling and the trust region shrinks. Such minima may be ghosts of orbits that exist at nearby values of the parameters \citep{zheng2025ghost}, or come from near recurrences whose ends are close in norm but far apart dynamically, perhaps on different sheets of a strongly folded attractor.

\begin{figure}[t]
\centering
\includegraphics[width=\textwidth]{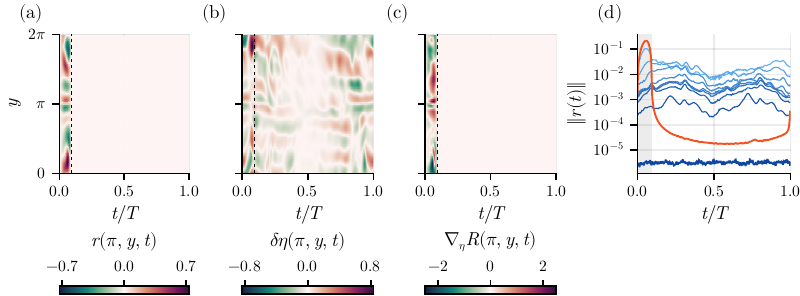}
\caption{The first Newton correction and the gradient in a search from a loop of period 42.9 that converges to RPO78. (a) The residual of the initial loop, (b) the correction of the first Newton iteration and (c) the gradient \eqref{eq:gradient} of $R$ on the initial loop, at $x = \pi$, against $t/T$ and $y$. (d) The norm of the residual at each instant, at the start (orange) and every third Newton iteration (light to dark), with the window of the weighted average shaded, its end dashed in (a--c).}
\label{fig:residualtime}
\end{figure}

The adjoint-based methods compute the gradient of $R$ directly. Integrating by parts in $s$, $x$ and $y$ in the inner product \eqref{eq:inner}, as in \citet{farazmand2016adjoint} and \citet{parker2022variational}, gives
\begin{equation}
\nabla_\eta R = \cJ_\eta^\dagger\, r = -\omega\,\partial_s r + c\,\partial_x r - \cL^\dagger(s)\,r ,
\label{eq:gradient}
\end{equation}
where, for Kolmogorov flow,
\begin{equation}
\cL^\dagger(s)\,r = \bu[\eta]\cdot\nabla r + \nabla^{-2}\big(\nabla^\perp r\cdot\nabla\eta\big) + \frac{1}{\Rey}\nabla^2 r
\label{eq:adjoint}
\end{equation}
is the adjoint of the linearised operator \eqref{eq:linearised}. Figure~\ref{fig:residualtime} shows how the gradient and the Newton correction differ, on a search from a loop of period 42.9 that converges to RPO78. The residual of the initial loop is largest in the window of the weighted average (figure~\ref{fig:residualtime}a). The first Newton correction spreads over the whole loop (figure~\ref{fig:residualtime}b), since the Newton system couples all instants, and lowers the residual at the junction while raising it elsewhere to about $10^{-2}$, after which the residual falls at every instant at a similar rate (figure~\ref{fig:residualtime}d). The gradient, local in time, is confined to the window (figure~\ref{fig:residualtime}c), so that a correction by descent reaches the rest of the loop only over many steps, which helps to explain why methods of descent converge slowly. A gradient, however, costs about one product with $\cJ$, far less than a hookstep, which builds a whole Krylov space. A first stage of descent, as is done with shooting \citep{page2024recurrent, page2025computation}, may thus still save time, and is left to future work.

\subsection{Wall time and memory}
\label{sec:cost}

The Newton iterations of the searches that converge grow only slowly with the period $T_0$ of the initial loop (figure~\ref{fig:scaling}a), but the Arnoldi steps per Newton iteration grow linearly with it beyond the shortest loops (figure~\ref{fig:scaling}b). The extra steps form the plateau of the GMRES histories (figure~\ref{fig:traces}d, h), on which GMRES resolves the eigenvalues of $\cJ\cB^{-1}$ near the origin one by one \citep[Lecture~35]{trefethen1997numerical}. They belong to the slow scales of the loop, on which $\cJ$ is small and the preconditioner is about $\sigma\mathsf{I}$, so that it leaves them near the origin. Their number grows with the period, as the plateau does, and the corresponding Ritz vectors have most of their energy at temporal frequencies below $\sigma$ (not shown).

Since the time of an Arnoldi step also grows in proportion to the period (appendix~\ref{app:speedup}), the wall time of the searches that converge grows unfavourably, about as the square of the period (figure~\ref{fig:scaling}c), and so does the memory of the Krylov space, whose vectors grow in number and in size. This suggests that further algorithmic steps are needed to bring this scaling to a more favourable order in the period. The same growth set the longest periods of our campaigns, about 200, and may contribute to the stalling of some of the longest searches, which fill their Krylov space before GMRES reaches its tolerance (figure~\ref{fig:scaling}b), although a search may still converge with a truncated Krylov space, if more slowly.

\section{Campaigns of searches}
\label{sec:searches}

We report the three campaigns (\S\ref{sec:protocol}), the seeds that converge (\S\ref{sec:outcome}) and the orbits found, focusing on the long ones (\S\ref{sec:orbits}).

\subsection{The campaigns}
\label{sec:protocol}

The searches ran in three campaigns as the method matured (table~\ref{tab:campaigns}), on $3\,530$ distinct seeds, some searched in more than one campaign. The seeds are near recurrences of \S\ref{sec:guesses}, of which we keep only the closest among those of the same simulation within 50 time units and 10\% in period of each other, which are taken to follow the same orbit. Several GPUs took the seeds in turn from classes of period ten time units wide, so that all periods were searched from the start, for $1\,515$ GPU hours in all, about 23 minutes per search. Late in the third campaign we also tried the hookstep at several radii in the same Krylov space, keeping the step of smallest residual, and restricted the seeds, first to periods above 20 and then mostly to periods above 50 whose dissipation stays below 0.16 of the laminar value, preferring the seeds of larger dissipation. The seeds searched are therefore not a uniform sample in period or in dissipation. A converged orbit counts as new unless one of the same kind was found before with the same mean dissipation, to a relative difference of $10^{-3}$, and a period in an integer ratio to its own, a loose tolerance because the searches in single precision fix the mean dissipation only to a few parts in $10^4$.

\begin{figure}[t]
\centering
\includegraphics[width=\textwidth]{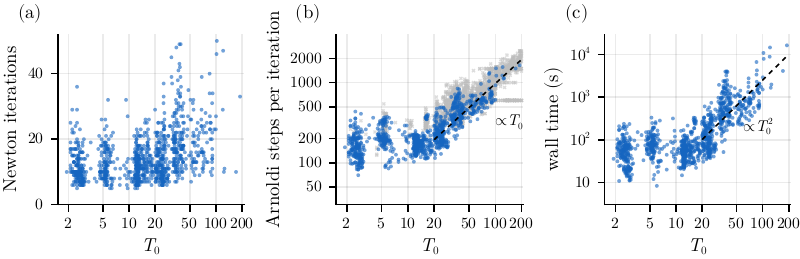}
\caption{The Newton iterations, Arnoldi steps and wall time of the searches against the period $T_0$ of the initial loop. (a) The Newton iterations of the searches that converge. (b) The median number of Arnoldi steps per Newton iteration, for the searches that converge (blue) and those that do not (grey crosses). (c) The wall time of the searches that converge. The dashed lines are proportional to $T_0$ in (b) and to $T_0^2$ in (c).}
\label{fig:scaling}
\end{figure}

\begin{table}
\begin{center}
\def~{\hphantom{0}}
\begin{tabular}{lccccrrr}
campaign & $N_s/T$ & precision & $\sigma$ & Krylov & searches & converged & solutions \\[3pt]
first                & 10      & double        & 2.5     & 600--1\,000   & 494     & 107 & 16 \\
second               & 4       & mixed, single & 1, 2.5  & 1\,500        & 537     & 188 & 52 \\
third                & 8, 6--7 & single        & 1       & 1\,500--3\,000 & 2\,931 & 736 & 140 \\[3pt]
all                  &         &               &         &               & 3\,962  & 1\,031 & 183 \\
\end{tabular}
\caption{The three campaigns, with the instants per unit time, the precision, the shift of the preconditioner, the largest Krylov space and the numbers of searches, converged searches and distinct solutions, which overlap between campaigns. Mixed precision is single precision down to $\lVert r\rVert = 10^{-5}$ and double precision after. The searches ran on NVIDIA L4, L40S, A100 and H200 GPUs, with 24, 48, 80 and 141~GB of memory.}
\label{tab:campaigns}
\end{center}
\end{table}

\subsection{Which seeds converge}
\label{sec:outcome}

\begin{figure}[t]
\centering
\includegraphics[width=\textwidth]{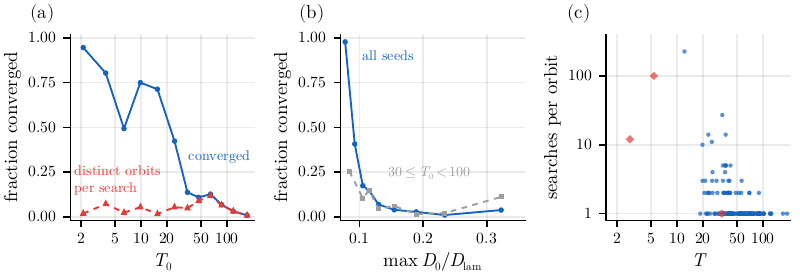}
\caption{The outcome of the 3\,962 searches. (a) The fraction of searches that converge, in bins of the period $T_0$ of the seed, and the distinct orbits reached per search (triangles). (b) The same fraction in bins of the largest dissipation along the seed, for all the seeds (blue) and for $30 \le T_0 < 100$ (grey), without bins of fewer than five searches. (c) The number of searches that converge to each periodic (diamonds) and relative periodic orbit (discs) against its period, without the travelling wave and the equilibrium, found 308 and 73 times.}
\label{fig:outcome}
\end{figure}

Of the $3\,962$ searches, $1\,031$ converged, to 183 distinct solutions, 178 of them relative periodic orbits. The longest, RPO178, of period 190.79, is more than twice as long as the longest of \citet{beck2026identifying}, of period 74.07. The fraction of searches that converge falls steeply with the period of the seed (figure~\ref{fig:outcome}a). This fall is commonly ascribed to the Floquet multipliers of long orbits, which grow exponentially with the period, but this growth need not make good seeds rarer (\S\ref{sec:guesses}). Part of the difficulty may lie in the large Krylov spaces of the longest loops (\S\ref{sec:cost}), although multiple shooting also fails on almost all the seeds of period above 100 that we tried (\S\ref{sec:shooting}). Another part may come from the bursts, which a longer seed is more likely to approach, since the fraction also falls with the largest dissipation along the seed, at similar periods (figure~\ref{fig:outcome}b). Over all periods, the quietest seeds mostly converge, to orbits in the bulk of the attractor, while only a few per cent of those that approach the bursts do so, mostly to long bursting orbits.

Most converged searches are duplicates. Two thirds of them, and almost all of those from seeds of period below 20, return to four short solutions of \citet{chandler2013invariant}, the equilibrium E1, the travelling wave T1, the periodic orbit P1 and the relative periodic orbit R19 (figure~\ref{fig:outcome}c), since E1 and T1 solve the space-time problem at any period and P1 and R19 at multiples of their own. The turbulent flow visits these solutions often \citep{chandler2013invariant}, so that near recurrences cluster around the periods of P1 and R19 (\S\ref{sec:guesses}) and multiple shooting returns to them as well (not shown). Most long orbits, by contrast, were found by a single search, since near recurrences of long period rarely return to the same orbit (\S\ref{sec:guesses}), as for the long orbits of low-dimensional systems \citep{lasagna2020sensitivity}. Above a period of about 40, almost every search that converges thus finds a new orbit (figure~\ref{fig:outcome}a).

\subsection{The orbits found}
\label{sec:orbits}

\begin{figure}[t]
\centering
\includegraphics[width=\textwidth]{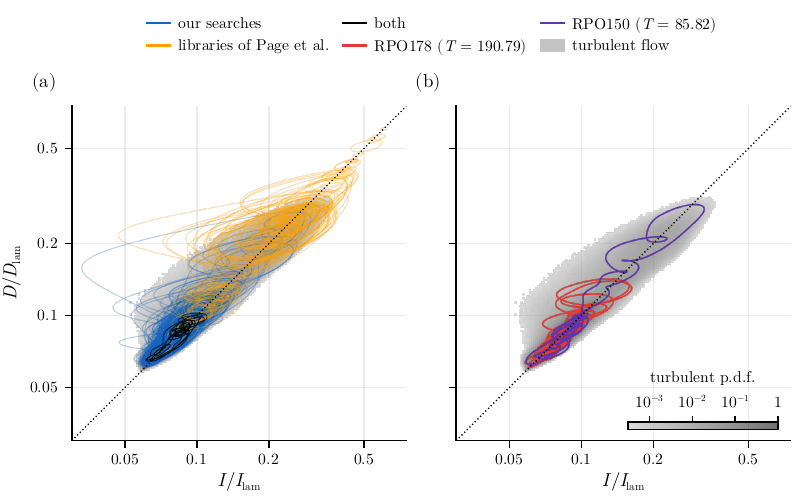}
\caption{The orbits at $\Rey = 40$ in the plane of the energy input and the dissipation, over the joint probability density of the turbulent flow (grey, on a logarithmic scale relative to its peak), with the diagonal $I = D$ (dotted). (a) The orbits found only by our searches (blue), only in the libraries of Page et al.\ (orange) and by both (black). (b) RPO178, the longest orbit (red), and RPO150, our long orbit of highest mean dissipation (purple).}
\label{fig:orbits}
\end{figure}

Our solutions, listed in table~S1 of the supplementary material, barely overlap with the short orbits of the two libraries of Page et al.\ \citep{page2024exact, page2024recurrent, cleary2025dynamical}, with only seven orbits in both, and occupy different ranges of the energy input and of the dissipation (figure~\ref{fig:orbits}a). These libraries are collected at invariant-sets.org, together with several thousand equilibria and travelling waves at this Reynolds number \citep{golden2026catalog}. Ours are mostly long and quiet, almost all with a mean dissipation below that of the turbulent flow. Nearly half of the orbits of \citet{chandler2013invariant} and few of those of \citet{page2024recurrent} are among ours, while only a handful of the solutions of \citet{redfern2024dynamically} are in either set. Our searches thus appear to miss the short bursting orbits that other methods find, as figure~\ref{fig:outcome}(b) suggests, although the form \eqref{eq:rpo}, which excludes orbits that close with a discrete symmetry, and the distance \eqref{eq:distance} may also play a part.

\begin{figure}[t]
\centering
\includegraphics[width=\textwidth]{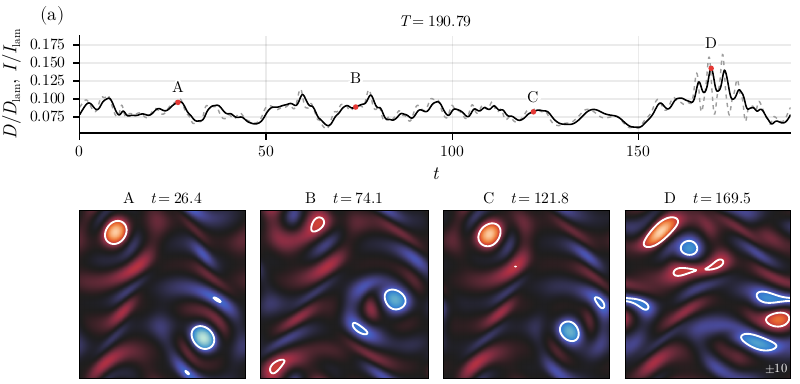}\\[4pt]
\includegraphics[width=\textwidth]{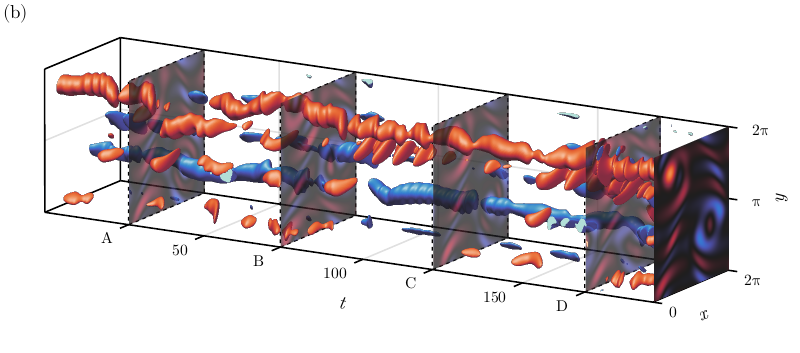}
\caption{RPO178, the longest orbit of our searches. (a) Its dissipation (solid) and energy input (dashed) along the period, and its vorticity at four instants A--D a quarter of the period apart (dots), the last at the largest dissipation, on a scale from $-10$ (blue) through $0$ (black) to $10$ (red), with the contours $\eta = \pm 5$ (white). (b) The isosurfaces $\eta = 5$ (red) and $\eta = -5$ (blue) over one period, in the frame that drifts with the orbit, lighter where the faces of the box cut a vortex, with the vorticity on the scale of (a) at the instants A--D on semi-transparent slices (dashed) and at $t = T$ on the end face. The orbit is translated by $\pi/2$ along $y$.}
\label{fig:gallery}
\end{figure}

Some long orbits pass through both the quiet states of the flow and the region of its bursts. RPO178 covers the bulk of the turbulent density in the plane of the energy input and the dissipation and rises once towards the bursts (figure~\ref{fig:orbits}b). Its flow is a pair of vortices of opposite sign, which this rise distorts but does not destroy (figure~\ref{fig:gallery}a). In RPO150, our long orbit of highest mean dissipation, the vortices instead dissolve into jets along the force and form again (not shown), as it climbs far into the region of the bursts (figure~\ref{fig:orbits}b). Long orbits thus need not stay quiet throughout their period, even when their mean dissipation is low. In space and time, the form in which the method represents and corrects an orbit, RPO178 appears as two vortex tubes that persist over the whole period and wander slowly across the domain, with small scales only near the peak of the dissipation (figure~\ref{fig:gallery}b).

\section{Verification by multiple shooting and a first comparison}
\label{sec:shooting}

We used multiple shooting to verify that the space-time solutions, most of which converged in single precision on a few instants per unit time, are orbits of the time stepper of the simulations. We also used this verification to take a first step towards a question often discussed in the literature, whether variational methods are more robust than shooting \citep{lan2004variational, farazmand2016adjoint, parker2022variational}. On this question, which is not the main aim of this paper, we only sought a first indication.

Multiple shooting splits the period into segments of about ten time units. Its unknowns are the states at the start of the segments, the period and the shift, and its equations require each segment, advanced by the time stepper, to end at the start of the next, and the last, translated by the shift, at the start of the first, with phase conditions in time and along $x$ \citep{ashtari2026computing}. The time stepper is that of the simulations, the third-order implicit-explicit scheme CB3R2R3e \citep{cavaglieri2015low} with steps of 0.01, which also advances the linearised equations, so that the Jacobian is that of the discrete flow map. We solve these equations in double precision by the Newton--Krylov hookstep method of \citet{viswanath2007recurrent}, with GMRES unpreconditioned, to a relative residual of $10^{-4}$ in at most 600 steps, and the trust region of \S\ref{sec:krylov}. A search stops when the residual of the shooting equations falls below $10^{-10}$, after 50 accepted Newton steps, or when this residual has not halved over the last 15 iterations.

Started from each space-time solution, with one of its instants at the start of each segment, multiple shooting converged all but one, most of them to within $10^{-10}$ in two Newton iterations. The space-time solutions thus lie well inside the basin of quadratic convergence of an orbit of the time stepper, at a distance consistent with their residual.

\subsection{A first step towards a comparison of the two methods}
\label{sec:comparison}

Robustness can be measured in several ways, all somewhat arbitrary. We take the probability that a search from a seed converges within the budget of its method, without matching the budgets, since the space-time searches have larger, preconditioned Krylov spaces. Multiple shooting searched a sample of seeds of the second and third campaigns, stratified in period and chosen by the outcome of their space-time searches, with about as many converged as failed below a period of 50 and mostly failed ones above.

\begin{table}
\begin{center}
\begin{tabular}{lcccc}
 & both & space-time only & shooting only & neither \\[3pt]
$0 < T_0 \le 20$     & 133 & 7  & 17 & 112 \\
$20 < T_0 \le 50$    & 122 & 23 & 9  & 136 \\
$50 < T_0 \le 100$   & 49  & 17 & 7  & 152 \\
$100 < T_0 \le 200$  & 0   & 2  & 1  & 197 \\
\end{tabular}
\caption{The seeds of a sample of the second and third campaigns, by band of period, that converge with both the space-time method and multiple shooting, with one only and with neither.}
\label{tab:shooting}
\end{center}
\end{table}

The two methods fail on largely the same seeds (table~\ref{tab:shooting}), which points to the seeds rather than to the method as the main cause of failure. Multiple shooting converges a few more seeds below a period of 20, where it is single shooting or nearly so, the space-time method a few more above, and above a period of 100 both converge almost none, but these differences rest on a few tens of seeds among two to three hundred in each band. With segments over which perturbations grow less than tenfold, our multiple shooting shares much of the robustness that variational methods owe to treating the whole orbit at once \citep{ashtari2026computing}, as expected from the close relation between the two (\S\ref{sec:introduction}). The seeds, finally, are near recurrences, and the two methods may differ more from rougher guesses \citep{azimi2022constructing, parker2022variational}, while whether one has a larger basin for long and strongly unstable orbits needs a larger study, on a sample chosen independently of the outcomes of the searches.

\begin{figure}[t]
\centering
\includegraphics[width=\textwidth]{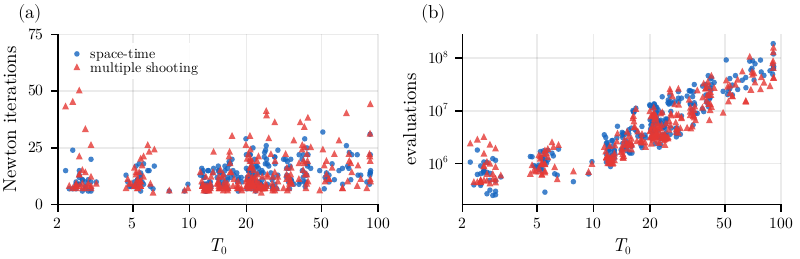}
\caption{The work of the two methods on the 263 seeds of the third campaign that both converge, for the space-time method (discs) and multiple shooting (triangles), against the period $T_0$ of the seed. (a) The Newton iterations. (b) The evaluations of the vector field or of its linearisation at one instant, over the instants of the products with $\cJ$ and of the inner steps of the preconditioner for the space-time method, and over the stages of the time steps for multiple shooting.}
\label{fig:shooting}
\end{figure}

Figure~\ref{fig:shooting} compares the work of the two methods on the seeds of the third campaign that both converge. They take similar numbers of Newton iterations and, interestingly, of evaluations of the vector field or of its linearisation at one instant (counted as in the caption). The space-time method takes about eight times as many Arnoldi steps (not shown), but each on far fewer instants, about eight per unit time against the 100 steps per unit time of the time stepper, since its grid need only resolve the orbit (\S\ref{sec:implementation}), while the time steps are limited by the Courant--Friedrichs--Lewy condition of the explicit terms. Per unit time, an Arnoldi step thus evaluates the linearised operator of order a hundred times, once for the product with $\cJ$ and once per inner step at each instant, against about six hundred evaluations for a Krylov step of multiple shooting, two at each of the three stages of every time step, so that the larger number of Arnoldi steps is roughly balanced. In our view the two methods are about equally robust from these seeds, within the uncertainty of the sample, and we discuss what may set them apart in \S\ref{sec:conclusions}.

\section{Long orbits and the statistics of the turbulent flow}
\label{sec:statistics}

We now turn to the main point of the paper, the properties of long orbits. One remark is in order. We examine individual orbits, which are not a sample of all the orbits of a given period but, as in our earlier study \citep{lasagna2020sensitivity}, those that the searches could converge from near recurrences, a selection strongly biased here towards quiet orbits (figures~\ref{fig:outcome} and \ref{fig:orbits}). These are, however, the orbits that searches find in practice, and the results below describe them, not all the admissible ones.

\subsection{The instability of the orbits}
\label{sec:floquet}

A long orbit that stands for the flow should be about as unstable as the flow, and we compare the Floquet exponents of the long orbits with the Lyapunov exponents of the turbulent flow. We computed the leading Floquet exponents by the matrix-free subspace iteration of \citet[][\S6.3.2]{lust1997thesis}, also used in appendix A of \citet{lasagna2020sensitivity}, on the flow map of the time stepper on which the orbits were converged by multiple shooting. It iterates a subspace of chosen dimension towards that of the leading multipliers, and keeps the projection of the monodromy matrix onto it as a product of small factors, corresponding to the action of the Jacobian of the flow map over segments of chosen length. The multipliers follow from the periodic Schur decomposition of these factors, without forming the product, which resolves them over many decades for long orbits, whereas an Arnoldi iteration on the monodromy matrix loses those whose ratio to the leading one falls below the machine precision. The Lyapunov exponents of table~\ref{tab:lyapunov}, denoted by $\Lambda_k$, were computed by the method of \citet{benettin1980lyapunov}. For P2, P1 and R19 of \citet{chandler2013invariant}, and for their P3, which our searches did not find but the libraries of Page et al.\ hold, the number of unstable exponents and the leading exponent agree with those of \citet{page2024recurrent} to the digits given there.

\begin{figure}[t]
\centering
\includegraphics[width=\textwidth]{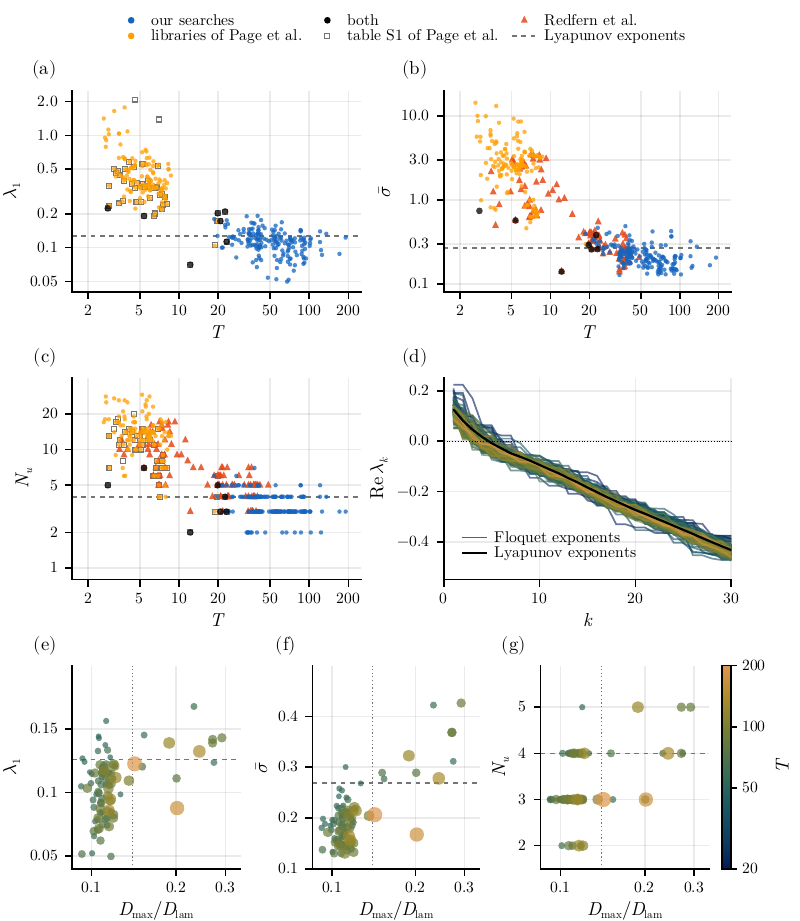}
\caption{The instability of the orbits against their period. (a) The leading Floquet exponent $\lambda_1$, (b) the sum $\bar\sigma = \sum_{\mathrm{Re}\,\lambda_k > 0}\mathrm{Re}\,\lambda_k$ of the unstable exponents and (c) their number $N_u$, without the two zero exponents, also, where available, for the orbits of the libraries of Page et al., of table~S1 in the supplementary material of \citet{page2024recurrent} and of \citet{redfern2024dynamically}, with the same quantities for the turbulent flow (dashed), averaged over the trajectories of table~\ref{tab:lyapunov}. (d) The leading Floquet exponents of our orbits against their index $k$, coloured by period, with the mean Lyapunov exponents, both without the two zero exponents. (e--g) The quantities of (a--c) for our orbits longer than 50 against the largest dissipation $D_\mathrm{max}$ along them, relative to the laminar dissipation, coloured and sized by period, with the value 0.14 above which an orbit bursts (dotted).}
\label{fig:floquet}
\end{figure}

\begin{table}
\begin{center}
\def~{\hphantom{0}}
\setlength{\tabcolsep}{2.5pt}
\small
\begin{tabular}{cccccccccc}
$\Lambda_1$ & $\Lambda_2$ & $\Lambda_3$ & $\Lambda_4$ & $\Lambda_7$ & $\Lambda_8$ & $\Lambda_9$ & $\Lambda_{10}$ & $\Lambda_{11}$ & $\Lambda_{12}$ \\[3pt]
0.126 & 0.081 & 0.045 & 0.017 & $-0.008$ & $-0.029$ & $-0.048$ & $-0.062$ & $-0.077$ & $-0.095$ \\
\end{tabular}
\caption{The leading Lyapunov exponents of the turbulent flow, averaged over four trajectories of 20\,000 time units, without the neutral exponents $\Lambda_5$ and $\Lambda_6$, below $10^{-3}$ in magnitude.}
\label{tab:lyapunov}
\end{center}
\end{table}

Figure~\ref{fig:floquet} shows that the long orbits are much less unstable than the short orbits of the libraries of Page et al., which have large leading exponents and often more than ten unstable ones, while the orbits of \citet{redfern2024dynamically}, of periods up to about 50, link the two sets. Between periods of 50 and 100 the sum of the unstable exponents of the long orbits is about a third below that of the positive Lyapunov exponents, and they mostly have three unstable exponents against four positive Lyapunov exponents (figure~\ref{fig:floquet}c and table~\ref{tab:lyapunov}), a number that varies from orbit to orbit, a sign that the flow is not uniformly hyperbolic, since on a uniformly hyperbolic attractor all orbits have the same number of unstable directions. Their leading Floquet exponents capture the Lyapunov spectrum quite well, perhaps slightly below it (figure~\ref{fig:floquet}d). The long orbits thus appear a little less unstable than the flow, which may reflect the under-representation of bursting seeds, since the flow is most unstable as it enters a burst \citep{farazmand2016dynamical}. Indeed, the long orbits that burst, whose largest dissipation exceeds 0.14 of the laminar value, are typically about as unstable as the flow in $\lambda_1$, $\bar\sigma$ and $N_u$, while the quiet ones lie below it (figure~\ref{fig:floquet}e--g).

These results can be explained by noting that both sets of exponents are averages of a local rate of growth. Along a trajectory $\eta(t)$, an orbit or the turbulent flow, a tangent perturbation obeys $\partial_t\delta\eta = \cL(t)\,\delta\eta$, with $\cL(t)$ the linearisation \eqref{eq:linearised} of $\cN$ about $\eta(t)$. Along a Floquet vector $\xi_k(t)$ of an orbit, or a covariant Lyapunov vector of the flow, denoted in the same way, a tangent vector grows or decays at the instantaneous rate
\begin{equation}
\kappa_k(t) = \frac{\big(\xi_k(t), \tfrac12\big(\cL(t) + \cL^\dagger(t)\big)\,\xi_k(t)\big)}{\big(\xi_k(t), \xi_k(t)\big)} ,
\label{eq:localexp}
\end{equation}
the local exponent \citep{ding2016estimating}, with $\cL^\dagger(t)$ the adjoint of $\cL(t)$, and
\begin{equation}
\mathrm{Re}\,\lambda_k = \frac{1}{T}\int_0^T \kappa_k(t)\,\dd t , \qquad
\Lambda_k = \lim_{t_1\to\infty}\frac{1}{t_1}\int_0^{t_1} \kappa_k(t)\,\dd t ,
\label{eq:floquetlyapunov}
\end{equation}
the first average over the period of the orbit and the second along a turbulent trajectory. A short orbit samples these rates only in the small region of the attractor that it visits, which would explain why the short orbits of the libraries, most of which spend their whole period in the tail of the dissipation, are so unstable. A long orbit that stays close to a turbulent segment (\S\ref{sec:moved}) visits the quiet phases and the bursts in proportions closer to those of the flow, so that its exponents, like its averages, may approach those of the flow as its period grows, as figure~\ref{fig:floquet} suggests.

\subsection{How far Newton's method moves the loop}
\label{sec:moved}

We now ask how close the orbits stay to the turbulent segments from which they were found, that is, how far the Newton iterations move the initial loops. Let $\eta_0(\tau)$, $0 \le \tau < T_0$, be the initial loop of the search that found the orbit $\eta(t)$, $0 \le t < T$, both in the frames that drift with them. The distance
\begin{equation}
d_0(t) = \min_{\tau,\,\ell}\frac{\lVert\mathcal{T}_\ell\,\eta_0(\tau) - \eta(t)\rVert}{\lVert\eta(t)\rVert}
\label{eq:seeddistance}
\end{equation}
is normalised as the recurrence distance \eqref{eq:distance}. If the loop follows the orbit, $d_0(t)$ stays small, the minimiser $\tau(t)$ advances at the rate $T_0/T$, close to one, and the minimiser $\ell(t)$ stays nearly constant, conditions used in studies of shadowing \citep{budanur2017relative, suri2020capturing, krygier2021exact}. We measure time in escape times $1/\bar\sigma$, the mean time over which trajectories leave a small neighbourhood of the orbit.

\begin{figure}[p]
\centering
\includegraphics[width=\textwidth]{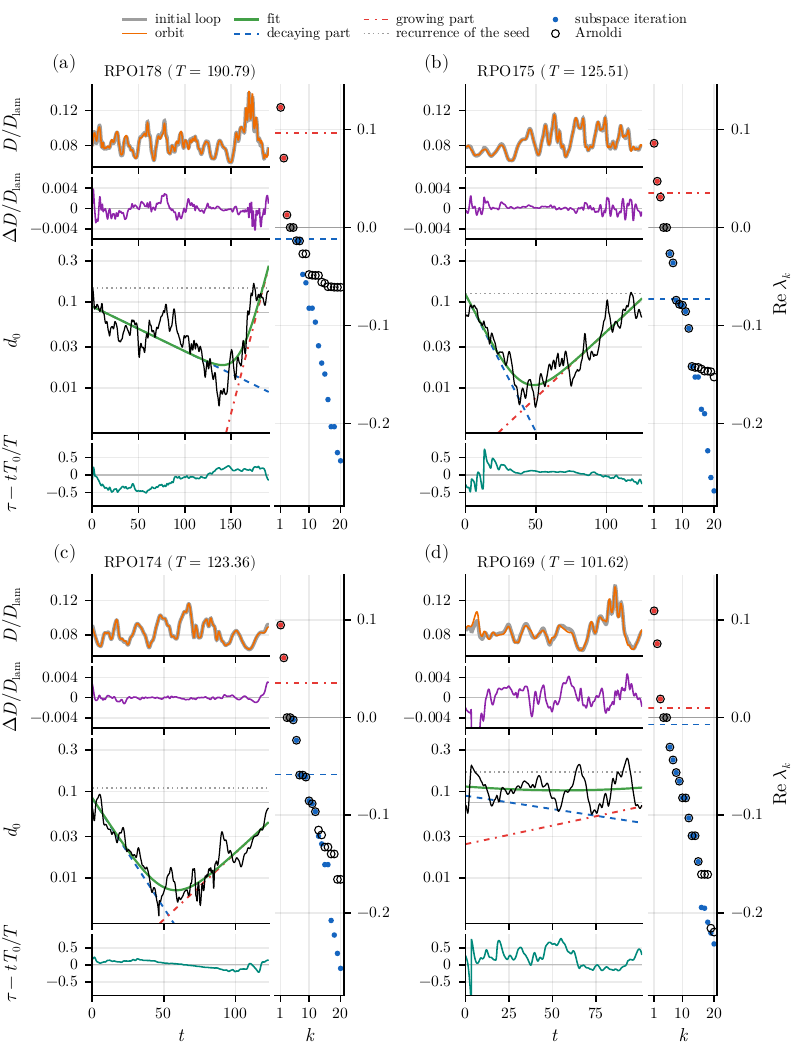}
\caption{Four long orbits against the initial loops of their searches, (a) RPO178, (b) RPO175, (c) RPO174 and (d) RPO169. In each block, from the top, the dissipation of the initial loop at $\tau(t)$ and of the orbit, their difference, the distance \eqref{eq:seeddistance} with the recurrence distance of the seed, the threshold 0.075 and the fit \eqref{eq:twoexp}, and $\tau(t) - tT_0/T$. On the right, the leading Floquet exponents of the orbit, up to twenty, from the subspace iteration of \citet{lust1997thesis} (dots, red if unstable, grey if neutral, blue if stable) and from the Arnoldi iteration on the monodromy matrix (circles), with the rates $\hat\lambda_s$ and $\hat\lambda_u$ of the decaying and the growing part of the fit.}
\label{fig:seeddistance}
\end{figure}

\begin{figure}[t]
\centering
\includegraphics[width=\textwidth]{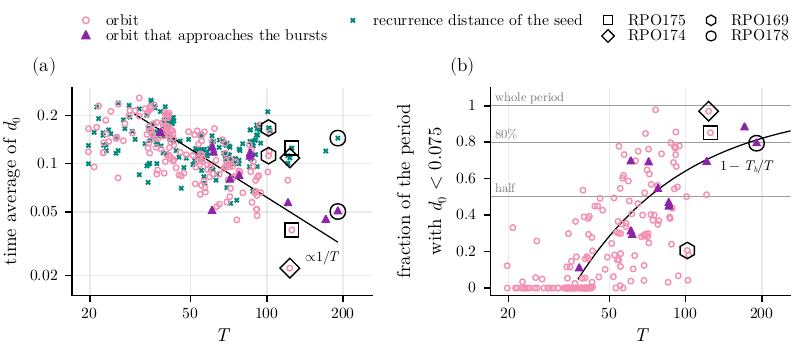}
\caption{How close the orbits of seed period $T_0 \ge 20$ stay to their initial loops, against their period. (a) The time average of the distance \eqref{eq:seeddistance}, with the recurrence distance of the seed (crosses) and a line proportional to $1/T$. (b) The fraction of the period with $d_0$ below the threshold 0.075, with the whole period, 80\% and half of it (grey) and the line $1 - T_b/T$ fitted by least squares. Quiet orbits are open circles and those that approach the region of the bursts filled triangles, and RPO178, RPO175, RPO174 and RPO169 of figure~\ref{fig:seeddistance} are ringed by a circle, a square, a diamond and a hexagon, in (a) also at the recurrence distance of their seeds.}
\label{fig:shadowcat}
\end{figure}

Figure~\ref{fig:seeddistance} shows four long orbits against the initial loops of their searches. In the first three (figure~\ref{fig:seeddistance}a--c) the distance stays small over most of the period, for tens of escape times, with $\tau(t)$ near $tT_0/T$ and a nearly constant $\ell(t)$ (not shown). It falls by more than a decade away from the junction, so that Newton's method does very little to these guesses. This closeness is largely built in, since each orbit is computed from the segment with which we compare it. That such segments exist may nevertheless seem surprising, since the probability of following a given orbit over its period falls as $\mathrm{e}^{-\bar\sigma T}$, and \citet{page2020searching} noted that recurrent flow analysis needs the flow to shadow an orbit over at least a whole period, which may become ever rarer as the Reynolds number grows. As discussed in \S\ref{sec:recurrences}, however, any of the exponentially many orbits of similar period would do, and what matters is the probability of a return (figure~\ref{fig:recurrences}d).

In these three orbits the distance falls from the start of the period and rises again towards its end, which we describe by the least-squares fit
\begin{equation}
d_0(t) \approx a\,\mathrm{e}^{\hat\lambda_s t} + b\,\mathrm{e}^{\hat\lambda_u (t - T)} ,
\label{eq:twoexp}
\end{equation}
with rates $\hat\lambda_s < 0 < \hat\lambda_u$, here commensurate with the leading Floquet exponents (figure~\ref{fig:seeddistance}, right), although not in general, as we discuss below. The fit works only where the orbit stays close to its loop. In the fourth orbit, whose distance falls below 0.075 for only a fifth of the period, there is no clear approach, and the fitted rates are close to zero (figure~\ref{fig:seeddistance}d).

The other orbits show similar trends (figure~\ref{fig:shadowcat}). The time average of $d_0$ falls about as $1/T$ beyond the shortest orbits (figure~\ref{fig:shadowcat}a), as expected from the boundary layers of approach and departure at the two ends of the period, which weigh less on the average as the period grows. Likewise, the fraction of the period spent within 0.075 of the loop (figure~\ref{fig:shadowcat}b) grows as $1 - T_b/T$, the boundary layers taking about $T_b = 35$ time units whatever the period, and covers nearly the whole period in a few orbits, tens of escape times for the longest, although it depends on this threshold. The recurrence distance of the seeds, by contrast, does not fall with the period (crosses in figure~\ref{fig:shadowcat}a), so that the Newton iterations move the loops of longer orbits less. This is almost paradoxical, since the searches of longer orbits start from loops closer to the orbit but need no fewer Newton iterations than those of shorter ones (figure~\ref{fig:scaling}a), as in our earlier searches in the Lorenz and Kuramoto--Sivashinsky systems \citep{lasagna2020sensitivity}. We speculate that the slow scales of the loop, whose number grows with the period (\S\ref{sec:cost}), must still be corrected, and that their nonlinearity limits the progress of each iteration.

Most distances fall and rise as \eqref{eq:twoexp} describes, but across the orbits found the fitted rates do not always lie close to a specific Floquet exponent (not shown), and in many orbits the distance decays more slowly than any stable exponent, or barely at all, so that the mechanism is probably more complex. The distance nevertheless falls and rises as if the segment approached the orbit along its stable manifold and left it along its unstable one, as in the picture of \citet{ding2016estimating} for the Kuramoto--Sivashinsky equation, where this happens within the span of the leading stable and unstable Floquet vectors of the orbit, its entangled physical modes \citep{yang2009hyperbolicity}, which span the tangent space of the inertial manifold, if one exists. Among the stable directions the least stable one dominates only after a time set by the gaps between the stable exponents and by how the initial difference projects onto their directions. Estimated from the Lyapunov exponents of the flow, close to those of the long orbits (\S\ref{sec:floquet}), the least stable direction gains a factor of ten over the next, from equal amplitudes, in a time $\ln 10/(\Lambda_7 - \Lambda_8)$ of about 110 time units (table~\ref{tab:lyapunov}), while in most of our orbits the approach is too brief for a single direction to dominate. A projection of the difference between the turbulent segment and the orbit onto the Floquet vectors, as \citet{ding2016estimating} did, together with the angles between these vectors, through which the non-normality of the linearised dynamics can make the distance grow or decay transiently at other rates, would test this picture directly, and we leave it for future work.

\subsection{Shadowing of long orbits by the turbulent flow}
\label{sec:shadowing}

\begin{figure}[t]
\centering
\includegraphics[width=\textwidth]{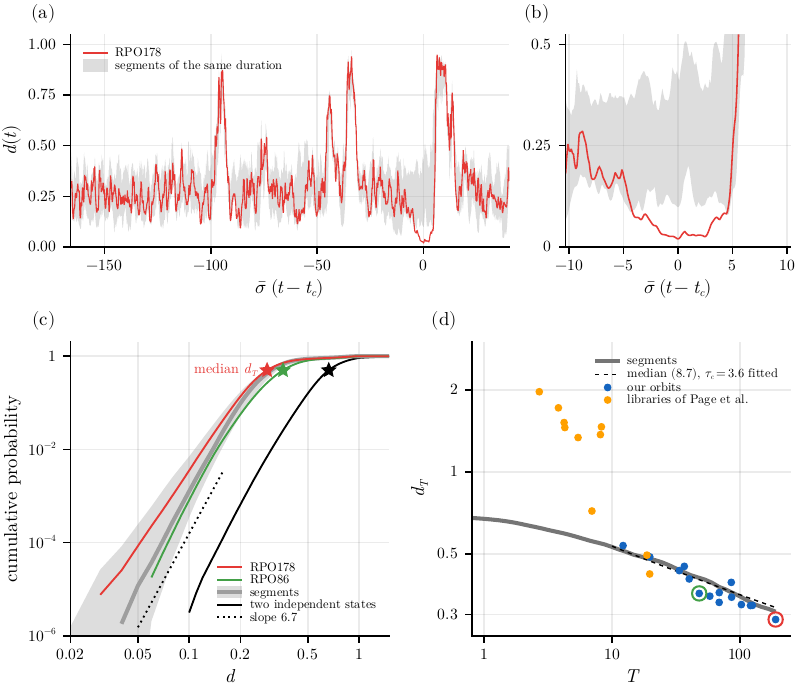}
\caption{The distance \eqref{eq:flowdistance} of turbulent trajectories from long orbits. (a) To RPO178, along the trajectory that comes closest to it, over 1\,000 time units about its closest approach at $t_c$, in escape times $1/\bar\sigma$ of the orbit, with the range of the distance to 32 turbulent segments of the same duration (grey). (b) The closest approach. (c) The cumulative distribution of the distance, the fraction of samples below $d$, to RPO178, over $1\,152$ trajectories of $2\,500$ time units (red, with a star at the median $d_T$), to RPO86 over 768 trajectories (green, with a star at its median), to the 32 turbulent segments of the same duration, over the 384 trajectories (grey, median and range), and between two independent states of the flow, the correlation integral $C(d)$ (black, with a star at its median), with a line of slope $D_2$. (d) The typical distance $d_T$ against the duration $T$, for turbulent segments (grey), from \eqref{eq:typical} with the measured $C(d)$ and $\tau_c = 3.6$ fitted for $T \ge 10$ (dashed), and for 25 orbits drawn across the periods, ours (blue) and from the libraries of Page et al.\ (orange), with RPO178 and RPO86 ringed. Here the states of the flow are those of the segments, one every time unit.}
\label{fig:closeness}
\end{figure}

Previous studies of shadowing have considered short orbits, asking whether turbulent trajectories come close to them and follow them \citep{krygier2021exact, crowley2022turbulence}, and how much of the time they spend near them \citep{suri2020capturing}. If a long orbit is to stand for the turbulent flow, and capture all its dynamically relevant states and events, the question is rather how much of a turbulent trajectory lies near the orbit, and how this grows with the period, since a longer orbit holds more states, spread over more of the attractor. We measure the distance of a turbulent trajectory from an orbit by
\begin{equation}
d(t) = \min_{\tau,\,\ell,\,g}\frac{\lVert\mathcal{T}_\ell\,g\,\eta_p(\tau) - \eta(t)\rVert}{\lVert\eta(t)\rVert} ,
\label{eq:flowdistance}
\end{equation}
where $\eta(t)$ is now a state of the flow, $\eta_p(\tau)$, $0 \le \tau < T$, a state of the orbit, and $g$ runs over the 16 symmetric copies given by the shift-reflect and the rotation (\S\ref{sec:kolmogorov}). We followed 384 trajectories of $2\,500$ time units each ($1\,152$ for the distribution of RPO178 and 768 for that of RPO86 in figure~\ref{fig:closeness}c), recorded after a transient of 500 time units from initial conditions other than those of our campaigns, and took as a yardstick the same distance from 32 turbulent segments of the period of each orbit, cut from other trajectories.

Figure~\ref{fig:closeness}(a) shows the distance of the trajectory that comes closest to RPO178, in escape times of the orbit, together with the range of its distance from the 32 turbulent segments of the same duration, and the two behave alike. Most of the time the distance stays near 0.25, the threshold of the near recurrences of \S\ref{sec:recurrences}, and it rises close to one during the bursts. Just before one of these bursts the flow comes within a few per cent of the orbit and follows it for about seven escape times (figure~\ref{fig:closeness}b), while the instant $\tau$ of the orbit advances with $t$ (not shown), a shadowing event in the sense of \citet{krygier2021exact}. Over all the trajectories, the cumulative distribution of the distance, the fraction of the time that the flow lies within $d$ of the orbit, is about the same as that from turbulent segments of the same duration (figure~\ref{fig:closeness}c, red and grey), so that, seen from the turbulent flow, RPO178 behaves as a turbulent segment of its length. Its median, about 0.29 (red star), is low compared with the median distance between two independent states of the flow, about 0.66 (black star), and lower than that of the shorter RPO86, of period 48 (green star).

A simple argument explains this behaviour, and predicts what to expect from orbits of longer period. Let $C(d)$ be the correlation integral of \S\ref{sec:recurrences}, the probability that two independent states of the flow lie within $d$ of each other, here over the symmetric copies. When a periodic orbit visits the attractor like the flow, a state sampled on it lies within $d_\ast$ of the turbulent state $\eta(t)$ with probability $C(d_\ast)$. If the orbit held $N$ independent states, all of them would lie farther than $d_\ast$ with probability $\bigl(1 - C(d_\ast)\bigr)^N$. But the states of such a trajectory are not independent, since the flow needs some time to move to a new region, and we count an orbit of period $T$ as $N = T/\tau_c$ independent states, with $\tau_c$ a decorrelation time, so that the flow lies farther than $d_\ast$ from it at a given instant with probability
\begin{equation}
\mathrm{Prob}[d > d_\ast] = \bigl(1 - C(d_\ast)\bigr)^{T/\tau_c} \approx \exp\!\left[-\frac{T}{\tau_c}\,C(d_\ast)\right] ,
\label{eq:closeness}
\end{equation}
where the last form holds when $C(d_\ast) \ll 1$. The probability that the turbulent flow does not come within $d_\ast$ of the orbit thus decays exponentially with the period, as the probability that a trajectory has not yet entered a small ball decays with time in many chaotic systems \citep{freitas2010hitting}. Two predictions follow. The typical distance $d_T$, the median of $d$, at which this probability is one half, satisfies
\begin{equation}
C(d_T) = 1 - 2^{-\tau_c/T} ,
\label{eq:typical}
\end{equation}
and since $C$ increases with $d$, the median distance falls as the period grows. At small distances, where $C(d_\ast) \ll \tau_c/T$, \eqref{eq:closeness} gives $\mathrm{Prob}[d \le d_\ast] \approx (T/\tau_c)\,C(d_\ast)$, so that the distribution of the distance has there the shape of the correlation integral, and rises in proportion to the period. Both predictions rest on counting the states of the orbit as independent, which holds only if the orbit keeps visiting new regions of the attractor, as a mixing trajectory does. An orbit that returns near states it has already visited, for instance one that follows a shorter orbit several times, holds fewer independent states than $T/\tau_c$, and the flow then comes close to it less often than \eqref{eq:closeness} predicts.

The decorrelation time $\tau_c$ depends on how it is measured. The autocorrelations of the dissipation, of the energy and of the coefficient of the forcing mode fall to $1/\mathrm{e}$ in about 11 time units, that of the vorticity field in about 30, and that of the velocity field more slowly still (not shown). These follow the slow large scales and the drift along $x$, and say little about where on the attractor the flow lies. We therefore fit $\tau_c$ to the median distance of turbulent segments of durations above 10, computing \eqref{eq:typical} by inverting the measured correlation integral (figure~\ref{fig:closeness}d). The fit gives $\tau_c \approx 3.6$ time units, of the same order as the 1.7 over which the distance \eqref{eq:flowdistance} between a state of the flow and its own future reaches a fraction $1 - 1/\mathrm{e}$ of that between two independent states, and far from the decorrelation times of the global quantities.

Figure~\ref{fig:closeness}(c, d) bears out both predictions. With this single fitted constant, \eqref{eq:typical} follows the median distance of the turbulent segments to within a few per cent over durations from 10 to 190 (figure~\ref{fig:closeness}d), and the 25 orbits drawn across the periods, RPO178 among them, follow the median of the segments of the same duration, except the short orbits of the libraries that burst, which sit far above it, in the region of the bursts, where the flow seldom goes. That these medians keep falling up to the longest periods suggests that the orbits, like the segments, keep visiting new regions of the attractor, as the count of independent states assumes. With the copies, the correlation integral has a slope $D_2 \approx 6.7$ between $d=0.15$ and $d=0.3$, an estimate of the correlation dimension close to the seven of \S\ref{sec:recurrences}, and the distributions of $d$ rise with about the same slope at small $d$, their tails roughly in proportion to the period, within a factor of about two, from RPO86 to RPO178 (figure~\ref{fig:closeness}c).

At a given instant, the turbulent flow thus comes within a given distance of a long orbit with a probability that grows with the period, since a longer orbit spans more of the attractor, as a turbulent segment does with its duration. This is favourable to the use of long orbits as proxies for the flow, but the growth is very slow, since its exponent is set by the dimension of the attractor. Where $C(d) \propto d^{D_2}$, \eqref{eq:typical} gives $d_T \propto (\tau_c/T)^{1/D_2}$ for long orbits, so that halving the typical distance requires a period about $2^{D_2} \approx 100$ times longer, a factor that depends somewhat on the norm in which $D_2$ is estimated and that grows exponentially with the dimension of the attractor, which itself grows with the Reynolds number \citep{cleary2025characterizing}. Even so, at this Reynolds number the flow lies half of the time within about $d=0.3$ of RPO178, whereas two independent states of the flow lie this close in less than one case in a hundred.

Closeness at one instant is not shadowing, in which the flow also follows the orbit for a while, and whether long orbits are shadowed more often than short ones, at some point of their period, is a separate question, which would need a longer analysis. Yet, since the tails of figure~\ref{fig:closeness}(c) rise in proportion to the period, as \eqref{eq:closeness} predicts, it is as if close approaches, and with them the chances of shadowing, became more frequent with the period.

\subsection{The statistics of the orbits}
\label{sec:kl}

The longest orbits found from near recurrences are close copies of turbulent segments over many escape times, and the longest is about as close to the turbulent flow as such a segment (\S\ref{sec:moved}, \S\ref{sec:shadowing}). Their statistics should then follow those of turbulent segments of the same duration, unless the closure of a segment selects orbits whose statistics differ systematically, as \citet{chandramoorthy2021probability} showed can happen for the trajectories of one-dimensional maps that shadow those of slightly perturbed maps.

\begin{figure}[p]
\centering
\includegraphics[width=0.97\textwidth]{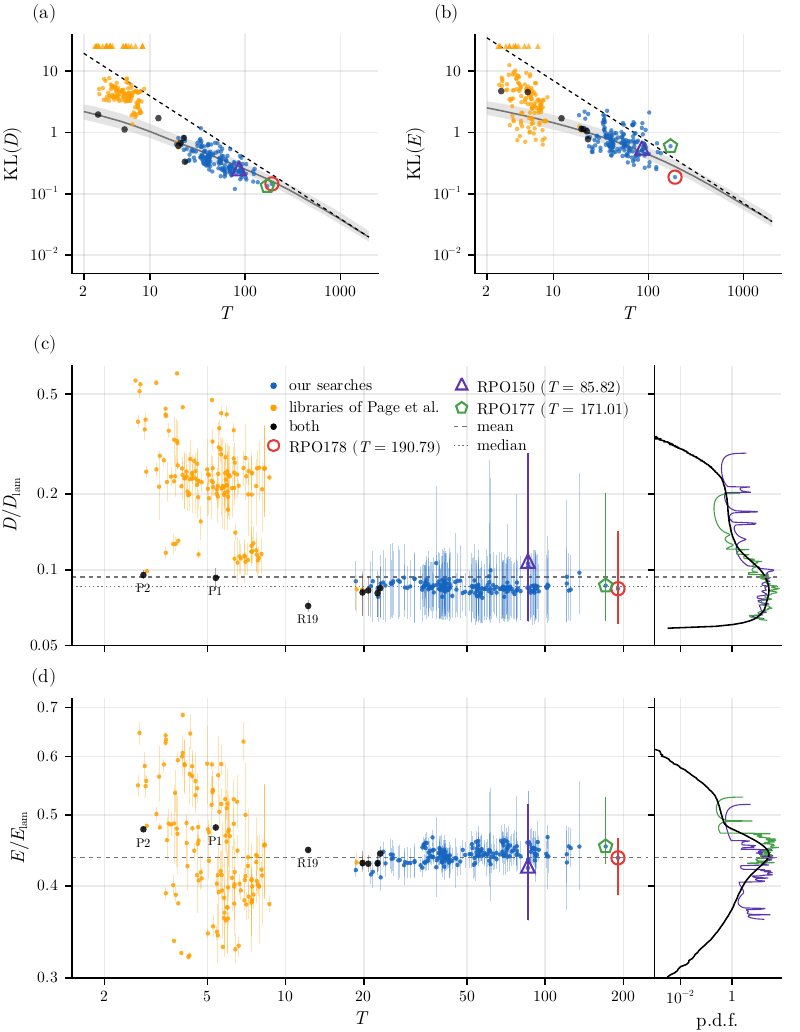}
\caption{The statistics of single orbits against those of the turbulent flow. (a, b) The Kullback--Leibler divergence \eqref{eq:kl} of each orbit from the flow against its period, $\mathrm{KL}(D)$ for the dissipation (a) and $\mathrm{KL}(E)$ for the kinetic energy (b), coloured as in (c), with the median (grey line) and the quartiles (grey band) of turbulent segments of duration $T$, and a line $\propto 1/T$ through the median at $T = 2\,000$ (dashed). Orbits beyond the largest value of the flow have an infinite divergence (triangles at the top). (c) The mean dissipation of each orbit against its period, with vertical lines spanning the dissipation along it, and the mean (dashed) and the median (dotted) of the flow. On the right, the probability density of the dissipation of the flow (black) and along RPO150 and RPO177. (d) The same for the kinetic energy, without the median, which nearly equals the mean, and without four short orbits of larger energy. RPO178, RPO150 and RPO177 are ringed by a circle, a triangle and a pentagon.}
\label{fig:statistics}
\end{figure}

We compare the orbits with the turbulent flow through the distributions of the dissipation and of the kinetic energy (figure~\ref{fig:statistics}), each relative to its laminar value. Along an orbit of period $T$ either quantity is a smooth periodic function $Q(t)$, and its probability density is the fraction of the period that it spends near each value $q$,
\begin{equation}
p(q) = \frac{1}{T}\int_0^T \delta\bigl(q - Q(t)\bigr)\,\dd t = \frac{1}{T}\sum_k \frac{1}{|\dot Q(t_k)|} ,
\label{eq:density}
\end{equation}
where the sum runs over the instants $t_k$ of the period at which $Q(t_k) = q$ \citep[eqs.~19.7 and 19.16]{chaosbook}. It is singular as $|q - Q_{\mathrm{e}}|^{-1/2}$ at every extremum $Q_{\mathrm{e}}$ of $Q(t)$, where $\dot Q$ vanishes. We evaluate its average over bins of width $\Delta q = 10^{-3}$,
\begin{equation}
p_i = \frac{1}{T\,\Delta q}\,\bigl|\{\,t \in [0, T) : q_i \le Q(t) < q_i + \Delta q\,\}\bigr| ,
\label{eq:binned}
\end{equation}
the total time that $Q(t)$ spends in the bin divided by $T\,\Delta q$, from the instants at which $Q(t)$, interpolated linearly between the steps of the time stepper on the orbit converged by multiple shooting (\S\ref{sec:shooting}), crosses the edges of the bin, so that the peaks at the extrema need no special treatment. The densities of the long orbits RPO150 and RPO177 spread over the bulk of the turbulent densities, with peaks at these extrema, and reach into the tail of the dissipation during their bursts (right of figure~\ref{fig:statistics}c, d).

We measure how far an orbit is from the turbulent flow by the Kullback--Leibler divergence of its density $p$ from the density $p_{\mathrm{DNS}}$ of the turbulent flow,
\begin{equation}
\mathrm{KL}(Q) = \int p(q)\,\log\frac{p(q)}{p_{\mathrm{DNS}}(q)}\,\dd q ,
\label{eq:kl}
\end{equation}
the measure that \citet{cleary2025dynamical} minimise to fit the weights of their library, and \citet{redfern2024dynamically} a loss of the same kind with the two densities exchanged. It vanishes when the two densities coincide and grows as the orbit spends its time where the turbulent flow seldom goes. Only the values of $Q$ that the orbit visits contribute, so that it is finite whenever the turbulent flow also visits them, whereas the reverse divergence is infinite unless the orbit reaches the whole range of the turbulent flow. We evaluate \eqref{eq:kl} on the same bins, with $p_{\mathrm{DNS}}$ from the samples of the turbulent flow, and compare each orbit with turbulent segments of the same duration, cut from about $2\times10^6$ time units of turbulent trajectories, whose divergence should decay as $1/T$ for long durations, as for the empirical distribution of independent samples \citep{paninski2003estimation}.

The long orbits of our searches follow the decrease of the divergence of these segments with the duration (figure~\ref{fig:statistics}a), which reaches the law $1/T$ only beyond about a thousand time units, once the segments visit the rare bins of the bursts many times, and those of period between 50 and 200 lie close to the median of the segments of the same duration. Of the two bursting orbits whose densities are drawn, RPO150 lies at the lower quartile and RPO177 below it. For the energy the long orbits scatter more widely (figure~\ref{fig:statistics}b), and RPO177 lies above the band, so that the energy is the more demanding test, as in the reconstructions of \citet{cleary2025dynamical} with weights fitted on the dissipation. The densities of the long orbits thus differ from that of the flow about as much as those of turbulent segments of the same duration. This must be qualified, since the divergence does not penalise an orbit for missing the tail of the turbulent density. A quiet orbit that stays near the peak of the density, as most of ours do (figure~\ref{fig:statistics}c), can therefore score as well as a turbulent segment that visits a burst, although it never reproduces the tail (right of figure~\ref{fig:statistics}c).

The mean dissipation and the mean energy of the long orbits lie near the median of the flow, while those of the short orbits scatter far more widely (figure~\ref{fig:statistics}c, d), and only some bursting orbits reach the mean dissipation, which the rare bursts raise above the median. If the long orbits keep representing turbulent segments of their duration as the period grows, their divergence should keep falling, eventually as $1/T$, and their means should approach those of the flow as $T^{-1/2}$, as those of turbulent segments do, two trends that remain to be confirmed. Testing this requires orbits about ten times longer, which would need an advance of the method or larger computational resources.

\section{Discussion and conclusions}
\label{sec:conclusions}

This work is part of a programme to compute the statistics of a turbulent flow, and their sensitivity to its parameters, on one or a few long unstable periodic orbits embedded in the attractor \citep{lasagna2018sensitivity, lasagna2020sensitivity}. On a periodic orbit the periodicity condition removes the exponential growth that the adjoint solution suffers on an open chaotic trajectory, and long orbits that capture the relevant dynamical events of the flow may stand as a proxy for how its attractor changes with the parameters, a step beyond the computation of averages from periodic orbits. Testing this hypothesis requires long orbits of a turbulent flow, which are hard to compute with the standard methods. We have therefore taken a step towards methods that can compute them, and examined, in two-dimensional Kolmogorov flow at $\Rey = 40$, whether the long orbits that can be found model the statistics of the flow.

The method is a preconditioned, matrix-free Newton--Krylov iteration on the whole loop, as advocated by \citet{lan2004variational}, implemented on GPUs. With the approximate factorisation of time-spectral methods as preconditioner, which retains advection, Newton's method on the loop appears to become practical for flows of this size, with a number of Arnoldi steps per Newton iteration that appears not to depend on the spatial resolution, at least at this Reynolds number. The classical trust-region approach \citep{dennis1983numerical}, known as the hookstep in the literature on shooting \citep{viswanath2007recurrent}, links the method to recent adjoint descent approaches \citep{azimi2022constructing}, since far from an orbit the hookstep descends the residual as these methods do, while near the orbit the iteration becomes an inexact Newton method that converges fast. A hookstep costs far more than a step of descent, but it also does more, since from the very first iterations it corrects the whole loop, while a gradient is local in time and may need many steps to propagate the correction through the period. Our searches did not need a separate stage of descent, but whether a hybrid method that first descends at a lower cost and then switches to hooksteps, as is done with shooting \citep{page2024recurrent, page2025computation}, would be faster remains to be verified.

A secondary question of this work was whether a method on the whole loop is more robust than multiple shooting. From near recurrences the two methods appear similar, since they fail on largely the same seeds and take similar numbers of Newton iterations and of evaluations of the vector field, although they may differ from rougher guesses \citep{azimi2022constructing, parker2022variational}. Comparing their computational costs would need a more careful analysis, on the same hardware and at the same precision, and with adjoint-based methods as well, which is beyond the scope of this work. The two methods also differ in what they require. A space-time method need only resolve the orbit in time, free of the stability limit of time stepping, and needs no time stepper, only the right-hand side, the linearised operator and a preconditioner, so that it may be simpler to implement once a preconditioner is available, although shooting can leverage existing time-stepping codes. Expanded in Fourier modes in time, the loop could also be projected onto resolvent or spectral proper orthogonal decomposition modes, as in searches for invariant solutions on a resolvent basis \citep{burton2026searching} and in the space-time reduced-order models of \citet{frame2027nonlinear} and \citet{li2025spacetime}, which could link the study of invariant solutions with model reduction.

Both methods were started from near recurrences of the turbulent flow, whose rate of occurrence we examined against the period. Recurrent flow analysis requires the flow to follow an orbit over a whole period \citep{page2020searching}, and the instability of long orbits might be expected to make this ever less likely as their period grows. The proliferation of long orbits, however, acts the other way, and the two effects may balance, as near recurrences of long period are about as frequent as those of moderate period (figure~\ref{fig:recurrences}e). From them the searches, mostly in single precision, converged 183 distinct solutions with periods up to 191, all but one verified in double precision by multiple shooting. The success of a search falls steeply with the largest dissipation along the seed and with its period, possibly because a longer recurrence is more likely to approach a burst and because the Krylov spaces of long loops grow large, while it depends little on how close the recurrence is, in the norm of the vorticity. Since multiple shooting fails on largely the same seeds, we speculate that, once the instability is tamed, the success of a search may depend on whether the two ends of its seed are dynamically connected, for instance on the same sheet of the folded attractor, which a small distance does not guarantee \citep{budanur2017relative, page2024exact}. Building initial guesses whose ends are dynamically connected, for instance by gluing shorter orbits \citep{budanur2017relative}, remains an open problem, and dynamic mode decomposition, which needs no near recurrence \citep{page2020searching}, would be worth testing on long orbits. The orbits that converge are mostly quiet, with a few bursting ones that capture part of the tail of the dissipation, and since they lie where the turbulent flow spends much of its time they may complement the short bursting orbits of existing libraries.

The orbits that we found are by necessity a selected sample of the unstable periodic orbits that exist, and they give partial answers to the questions raised in the introduction. A single long orbit appears to be a better proxy for the statistical properties of the turbulent flow than a single short one. Its mean dissipation and mean energy lie close to the medians of the flow, a necessary condition for its use, and the distribution of the dissipation along it is comparable with that of a turbulent segment of the same duration, that of the energy less so. This may be expected, since Newton's method moves these loops little, so that the longest orbits stay close to the segments from which they were found over most of their period, and their statistics appear to be those of a turbulent segment of their own length. The agreement does not reach the tail, since most orbits miss the rare bursts that raise the mean dissipation of the flow above its median. The missing bursts may also show in the instability of the orbits. The leading Floquet exponents of the long orbits follow the Lyapunov spectrum of the flow, perhaps slightly below it, as may be expected of orbits quieter than the flow, which is most unstable as it enters a burst. Seen from the turbulent flow, a long orbit also behaves as a turbulent segment of its length. A simple argument, with one fitted decorrelation time, predicts how its typical distance from the flow falls with the period, about as $T^{-1/D_2}$ where the correlation integral follows a power law, with $D_2$ the correlation dimension of the attractor. In this scaling, akin to the curse of dimensionality, the period needed to bring the flow within a given distance of the orbit grows as that distance to the power $-D_2$, that is, exponentially with the dimension of the attractor, which grows with the Reynolds number \citep{cleary2025characterizing}, so that a single orbit of given period may represent more strongly turbulent flows less closely. At this Reynolds number, however, the flow already lies half of the time within a small distance of our longest orbit, one within which two independent states of the flow lie in less than one case in a hundred. The tails of the distribution of this distance also rise in proportion to the period, as if close approaches, and with them the chances of shadowing, became more frequent along longer orbits, although shadowing itself remains to be measured.

The work has limitations, and several steps should follow. Above all, our orbits, of periods up to about forty escape times, are not yet long enough to show how their statistics evolve as the period grows, but they are a step towards such orbits. The work also considers a single weakly chaotic flow, at $\Rey = 40$ in a small doubly periodic domain. The method must still be extended to flows bounded by walls, where treating the pressure is an issue \citep{ashtari2023identifying}, either as a variable of its own or implicitly through a Poisson equation, in the residual as in the preconditioner. Higher Reynolds numbers, larger domains and three-dimensional flows may also need the loop and its Krylov space to be distributed over several GPUs. The inner steps of the preconditioner take most of the time spent on the operators. The preconditioner also leaves the eigenvalues of the slow scales of the loop near the origin, so that the Arnoldi steps grow linearly with the period, and the wall time and the memory grow about as its square. Preconditioners that also capture these slow scales, at a cost that grows only linearly with the period, could make longer orbits accessible, for instance by projecting the variational equation onto the unstable and nearly neutral directions, which keeps the Newton systems small \citep{wang2023reduced}. Besides this practical interest, it may be important to understand what these slow scales are and what role these nearly neutral modes play in the dynamics. Long orbits also give access to the structure of the tangent space of the flow, and the Floquet vectors of many orbits could estimate the dimension of the inertial manifold, if one exists, as for the Kuramoto--Sivashinsky equation \citep{ding2016estimating}.

Finally, the sensitivity of turbulent statistics to parameters remains the goal of this work. With long orbits at hand, the adjoint method for periodic orbits \citep{lasagna2018sensitivity} could now be tried on a turbulent flow, and compared with the change of the long-time averages between simulations at nearby parameter values. A caveat is that the sensitivity of an orbit diverges at a bifurcation point. In low-dimensional systems this produces power-law tails in the density of the sensitivities, which decays as their inverse cube, set by the normal form of the bifurcation, so that the sensitivity of a single orbit may occasionally be far from the typical value \citep{lasagna2020sensitivity}. This may also matter in Kolmogorov flow, whose branches of orbits fold repeatedly as the Reynolds number grows \citep{cleary2025dynamical}. Most of our long orbits have no Floquet multiplier close to one, apart from the two of the symmetries, but a few have one within about 10\% of it (not shown) and may lie near such a bifurcation, so that how long orbits depend on the parameters, and whether they survive bifurcations, matters for this goal. At higher Reynolds numbers, where by the argument of \S\ref{sec:shadowing} an orbit of given period may represent the flow less closely, one would also need to understand how the sensitivity of such an orbit, the proxy, compares with that of the attractor it represents, an object of ever higher dimension. Whether a single long orbit can capture how the statistics of a turbulent flow change with the parameters is the question we would most like to answer.
\begin{bmhead}[Acknowledgements.]
This work was developed during a sabbatical in the autumn of 2026 at the Emergent Complexity in Physical Systems Laboratory of the \'Ecole Polytechnique F\'ed\'erale de Lausanne (EPFL). D.L. is grateful to Tobias Schneider for the financial support of the visit and for a calm and productive environment, without which this work would not have been possible, and to the members of the laboratory for their welcome and for many useful conversations. The authors acknowledge the use of the IRIDIS High Performance Computing Facility, and associated support services at the University of Southampton, in the completion of this work.
\end{bmhead}

\begin{bmhead}[Funding.]
This work was funded by a Leverhulme Trust Research Project Grant (RPG-2024-368).
\end{bmhead}

\begin{bmhead}[Declaration of interests.]
The authors report no conflict of interest.
\end{bmhead}

\begin{bmhead}[Declaration of generative AI use.]
Generative AI was used throughout this work, through Claude (Anthropic, Claude Opus 5.5, through Claude Code). The interaction was a productive collaboration, in which the tool was essential to implement and test numerical ideas rapidly, many of which were then abandoned. The codes on which this paper rests were developed in this way, through a careful analysis of the generated code and the crafting of physically relevant unit and integration tests, as is standard practice for scientific software. Our role was to set the direction of the work, and to guide, filter and select what may be useful to the community, an experience that echoes the view of computers as `Monte Carlo' theory machines of \citet{jimenez2020computers}, which pose many questions of which only some are later found useful. In our experience the creative part of research lies increasingly in providing this direction. An initial draft of the text was also produced with the same tool, and was then edited into its final form by the authors through a long process of human supervision and revision, and the plotting scripts of all the figures were also written with it. The authors take full responsibility for the content of the paper.
\end{bmhead}

\begin{bmhead}[Data availability statement.]
\sloppy The codes used in this work are available at \url{https://github.com/Davide-Lasagna-s-Lab}. The catalogue of our solutions (\S\ref{sec:orbits}) is available at invariant-sets.org.
\end{bmhead}

\begin{bmhead}[Author ORCIDs.\\]
D. Lasagna, \href{https://orcid.org/0000-0002-6501-6041}{https://orcid.org/0000-0002-6501-6041}; \\
Thomas Burton \href{https://orcid.org/0000-0001-7998-2278}{https://orcid.org/0000-0001-7998-2278};
\end{bmhead}

\appendix

\section{Wall time of the operators on CPUs and GPUs}
\label{app:speedup}

\begin{figure}[t]
\centering
\includegraphics[width=\textwidth]{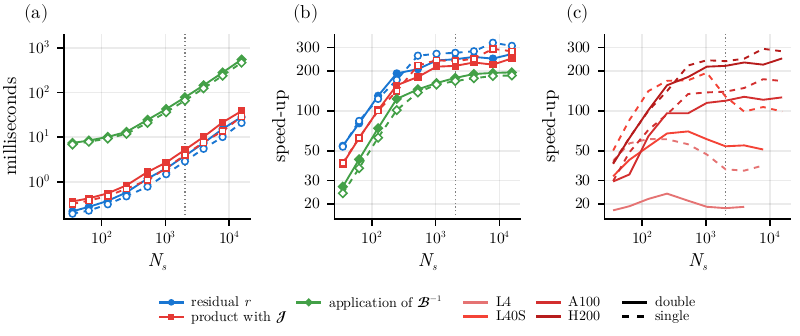}
\caption{The performance of the operators on GPUs, for Kolmogorov flow at $K = 33$, against the number of instants $N_s$, for the residual, the product with $\cJ$ and an application of $\cB^{-1}$ with the settings of the searches, in double (solid) and single precision (dashed). (a) The wall time on an NVIDIA H200 GPU. (b) Its speed-up over one core of an AMD EPYC 7502 processor. (c) The speed-up of the product with $\cJ$ on NVIDIA L4, L40S, A100 and H200 GPUs. The dotted line marks the largest loop of the searches.}
\label{fig:speedup}
\end{figure}

Figure~\ref{fig:speedup} compares the wall time of the operators of Kolmogorov flow at $K = 33$ on one CPU core and on GPUs. On the NVIDIA H200 GPU a product with $\cJ$ takes from a fraction of a millisecond to a few tens of milliseconds (figure~\ref{fig:speedup}a). The speed-up over the CPU core grows with the number of instants, from a few tens to a few hundred, until the device is fully occupied (figure~\ref{fig:speedup}b). The residual and the product with $\cJ$, which consist of transforms and element-wise operations alone, gain the most. The application of the preconditioner gains somewhat less, because at every inner step it exchanges $N_s$ numbers with the host. With ten inner steps it thus takes as long as 13 to 21 products with $\cJ$ on the H200, against about a dozen on the CPU core. Across the four GPUs (figure~\ref{fig:speedup}c), the speed-up of the product with $\cJ$ in double precision saturates at a level roughly proportional to their memory bandwidth, which limits such operations, and is an order of magnitude higher on the H200 than on the L4. In single precision the L4 and the L40S reach two to three times their speed-up in double precision, and the H200 only slightly more.

\clearpage

\bibliographystyle{jfm}
\bibliography{refs}

\end{document}